\documentclass[a4paper,11pt,fleqn]{article} 

\usepackage[centertags]{amsmath}
\usepackage{amssymb,bm}
\usepackage{amsfonts}

\usepackage{graphicx}
\usepackage{color}
\usepackage{placeins}
\usepackage[square]{natbib}

\newcommand{\dt}{\delta t}

\newcommand{\DT}{\Delta T}

\newcommand{\E}[1]{E\!\left[ #1 \right]}

\newcommand{\dn}{\delta n}
\newcommand{\cdf}{\operatorname{cdf}}
\renewcommand{\tt}{\sigma_{\dn}}
\newcommand{\ttMC}{\sigma_{\dn, \text{MC}}}
\newcommand{\ttEmp}{\sigma_{\dn, \text{emp}}}

\newcommand{\MC}{\text{MC}}

\newcommand{\nMC}{n_{\MC}}  
\newcommand{\NMod}{N_\text{mod}} 
\newcommand{\Ntrailing}{N_\text{trailing}}

\newlength{\figwidth}

\begin{document}


\begin{center}
 { \bf \huge Tile test for back-testing risk evaluation} 

\vspace{5ex}
{\bf\large 
	Gilles Zumbach
}\\[2ex]
\parbox{0.4\textwidth}{\renewcommand{\baselinestretch}{1.0}\normalsize
Edgelab\\
Avenue de la Rasude 5\\
1006 Lausanne\\
Switzerland
}
\\[3ex]
\parbox{0.4\textwidth}{\renewcommand{\baselinestretch}{1.0}\normalsize
gilles.zumbach@bluewin.ch\\

\vspace*{4ex}
\today
}


\begin{abstract}
A new test for measuring the accuracy of financial market risk estimations is introduced. 
It is based on the probability integral transform (PIT) of the \textit{ex post} realized returns using the \textit{ex ante} probability distributions underlying the risk estimation.
If the forecast is correct, the result of the PIT, that we called probtile, should be an iid random variable with a uniform distribution.
The new test measures the variance of the number of probtiles in a tiling over the whole sample. 
Using different tilings allow to check the dynamic and the distributional aspect of risk methodologies.
The new test is very powerful, and new benchmarks need to be introduced to take into account subtle mean reversion effects induced by some risk estimations.
The test is applied on 2 data sets for risk horizons of 1 and 10 days.
The results show unambiguously the importance of capturing correctly the dynamic of the financial market, and exclude some broadly used risk methodologies.

\end{abstract}
\end{center}
\vspace{2ex}


\newpage

\section{Introduction}
The evaluation of financial risk is an important part of any investment activity, and the regulators are imposing ever stricter rules over all participants.
The overall financial risk can be classified along several categories, and this paper concerns market risk (together with credit risk).
It is a major component of the financial risk, and it can be quantified efficiently.
In this context, a difficult issue is the validation of the quantitative figures provided by a risk evaluation.

A validation of a risk computation is done using back-testing.
The core idea of a back-test is fairly simple: using a long sample of historical data, evaluate the risk at different dates, and check over the following days that the risk has been correctly computed.
In details, the scheme is more complex.
The primary outcome of a risk evaluation is a probability distribution for the returns over a selected risk horizon $\DT$.
From this distribution, the usual risk measures can be computed like standard deviation, value-at-risk (VaR) or expected shortfall (ES, also called CVaR).
With the new information obtained each day, this distribution is changing in time.
As attested by the large body of literature on the volatility dynamics and the heteroskedasticity, the dynamics is quantitatively large and the volatility can change easily by a factor 5 over different periods (quiet, volatile or crisis).
Hence, we are dealing with a clearly non stationary system, with widely changing distributions.

The core of the back-test algorithm is better explained at a one-day risk horizon.
Each day, a forecast of the probability distribution of the returns is made, an \textit{ex ante} evaluation.
The next day, the actual return realized by the market becomes available, an \textit{ex post} value.
This single return should be drawn from the distribution computed on the previous day.
The goal of the back-test is to assess if the sequence of \textit{ex post} draws is coming from the sequence of \textit{ex ante} probability distributions. 
This is clearly not an easy problem, quite different from the usual test with repeated draws from a fixed distribution.

A simple possible test is provided by VaR: since it is defined as the $\alpha$ quantile of the loss distribution, the fraction of exceedances should be $1-\alpha$. 
This reduces the back-test to a simple counting exercise.
The weaknesses of this simple test are to check only one level for $\alpha$, and to ignore the dynamical aspect of the risk evaluation.
Better tests using the whole distribution of exceedances and/or the independence of the exceedances where proposed and improved in many contributions \citep{Kupiec.1995}, \citep{CrnkovicDrachman.1996}, \citep{DieboldGuntherTay.1998}, \citep{Barbachan.2006}, \citep{Christoffersen.1998}, \citep{ChristoffersenPelletier.2004}, see also the reviews by \citep{Campbell.2006}, \citep{Christoffersen.2010} and the master thesis of \citep{Roccioletti.2016}.

Better tests can be built using a probability integral transform (PIT): given a probability distribution and a return, compute the cumulative probability associated to the return. 
We named  ``probtile'' this realized $p$-value, since it corresponds to the cumulative probability of a quantile.
If the forecast for the pdf is correct, the probtiles should be iid with a uniform distribution. 
After the PIT transformation, the backtest problem becomes a standard statistical exercise for which numerous tests can be used, say for example a Kolmogorov-Smirnof test.

This testing strategy is known for a long time by statisticians, going back at least to \citep{Rosenblatt.1952}, and has been rediscovered several times in finance \citep{Christoffersen.1998, Zumbach.backtesting}.
Its main advantage is to test the whole distribution, and tests can be constructed to emphasize the tails which are the focus point for risk evaluation.
The dynamical aspect of the risk evaluation can also be tested, for example using lagged correlations of the probtiles.

An interesting recent development concerns scoring functions, elicitability and the direct comparison of risk methodologies.
The set-up is to have a fixed distribution for a random variable, playing the role of the returns.
The idea is to have some functions that attribute a score to the set (confidence level $\alpha$, tentative risk measure(s) , realized return).
For some selected scoring functions, the minimum of the expectation of the scoring function is given by the actual risk measure, say $\text{VaR}_\alpha$.
A risk measure for which such a scoring function exists is called elicitable.
An interesting property of the scoring function is to allow for a direct comparison of risk methodologies, essentially using the expected difference of the scores for the respective methodologies (see e.g. \citep{NoldeZiegel.2017} and references therein).
With this approach, various risk methodologies can be compared pairwise in order to decide which one is better.
Yet, the application of this approach to empirical data requires the distribution to be stationary, a hypothesis which is invalid in finance.
Given this limitation, we have to use an absolute approach, where each risk methodology is assessed independently for its qualities.

So far, most of the literature about financial risk is to check the distributional aspect of the forecast, and the dynamical aspect of risk evaluation has been much less discussed.
We find this bias quite odd, since having the dynamic right is quantitatively more important for actual decision-making than having the asymptotic distribution right.
We will first present a simple illustrative example for one stock, showing visually the key role of the dynamic. 
Then, we introduce the tile test that gives a joint test for the dynamics and the distribution.

\begin{figure}[ht]
	\centering
	\includegraphics[width=1.0\linewidth]{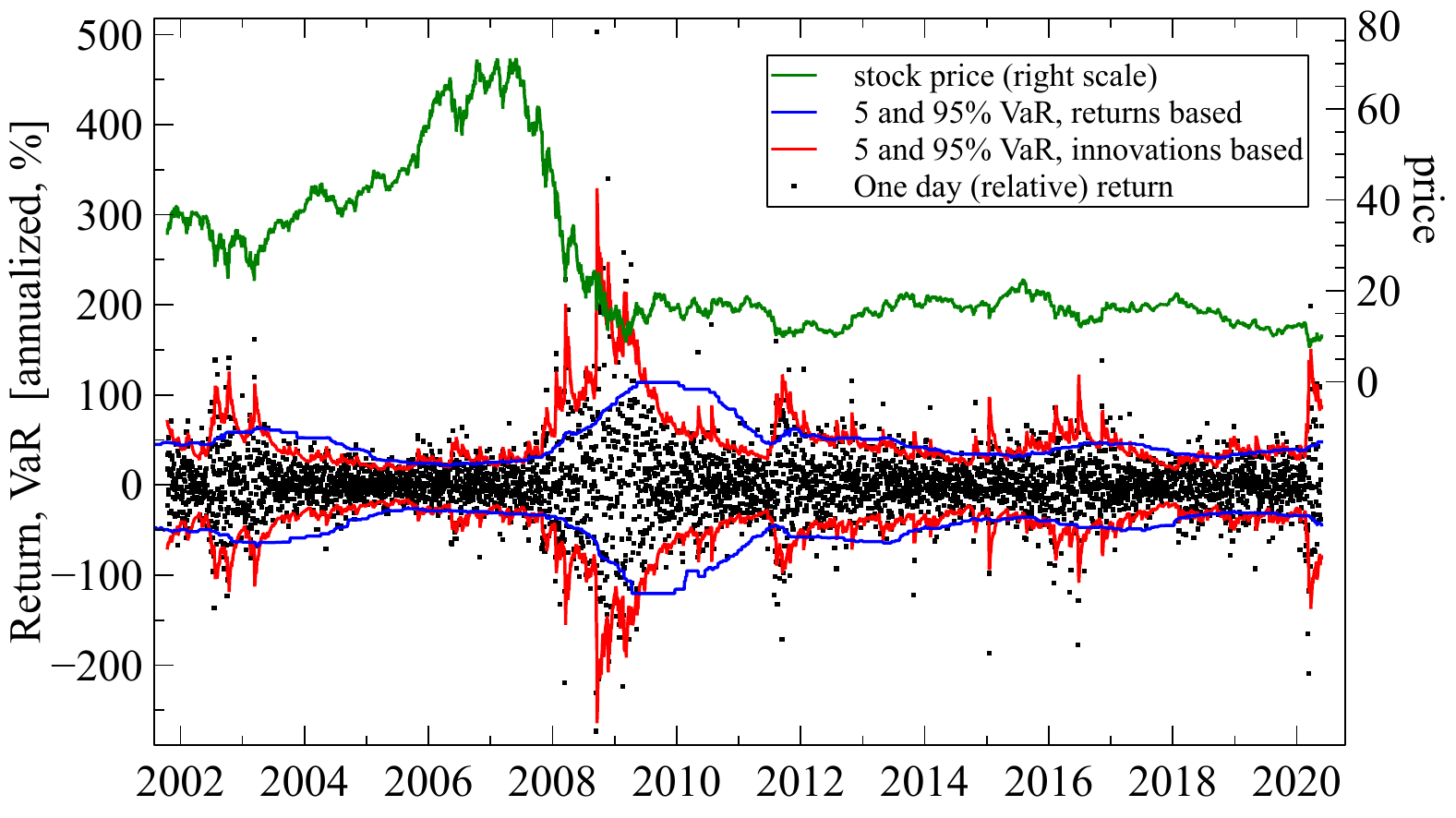}

	\caption{The price for the UBS stock (green line, right axis), the daily returns (black square) and the 5 and 95\% VaR, computed with the historical return methodology (blue line) and the historical innovation methodology (red line).}
	\label{fig:UBS_returns}
\end{figure}
The example uses the UBS stock price (a major Swiss bank), with a quite eventful history over the last 20 years related to the major crises and to events specific to UBS.
This choice provides for spectacular figures, but all stocks, indexes or FX show similar properties.
The figure~\ref{fig:UBS_returns} shows the UBS prices, the daily returns, and the 5 and 95\% VaR computed with two methodologies.
The daily returns are marked with black squares, and clearly the width of the return distribution changes with time.
Clear periods can be seen where the distribution is narrow or wide, a property called heteroskedasticity, namely the variance (skedasticity) of the return time series is not constant in time (hetero).
Accordingly, the risk is changing, and this should be reflected on the risk measures.

On Fig.~\ref{fig:UBS_returns}, the blue line corresponds to the widely used historical return methodology, namely the distribution for the next return is spanned by the daily returns in a moving 2 years window.
The red line is based on a long-memory ARCH process, with a historical distribution for the innovations  (details about the risk methodologies are given below in Sec.~\ref{sec:riskMethodologies}).
The dynamics is quite different between both models, with a slow (fast) adaptation to the market conditions for the blue (red) line.
If the risk computation is correct at the 5 and 95\% level, 5\% of the daily returns should be below and above the VaR lines.
This property should be verified not only asymptotically on a long sample, but at all times.
With naked eyes, we see that the red line is about right, but the blue line shows clear periods with too many or too little exceedances.
The difference can be large, see for example the covid crisis on the right of the graph.
Let us emphasize that the blue line can be wrong during extended periods, yet be correct on a long enough sample (i.e. asymptotically). 
This figure gives the gist of this paper, namely that the dynamics of the exceedances, or of the probtiles, should be tested thoroughly.

\begin{figure}[ht]
	\centering
	\includegraphics[width=1.0\linewidth]{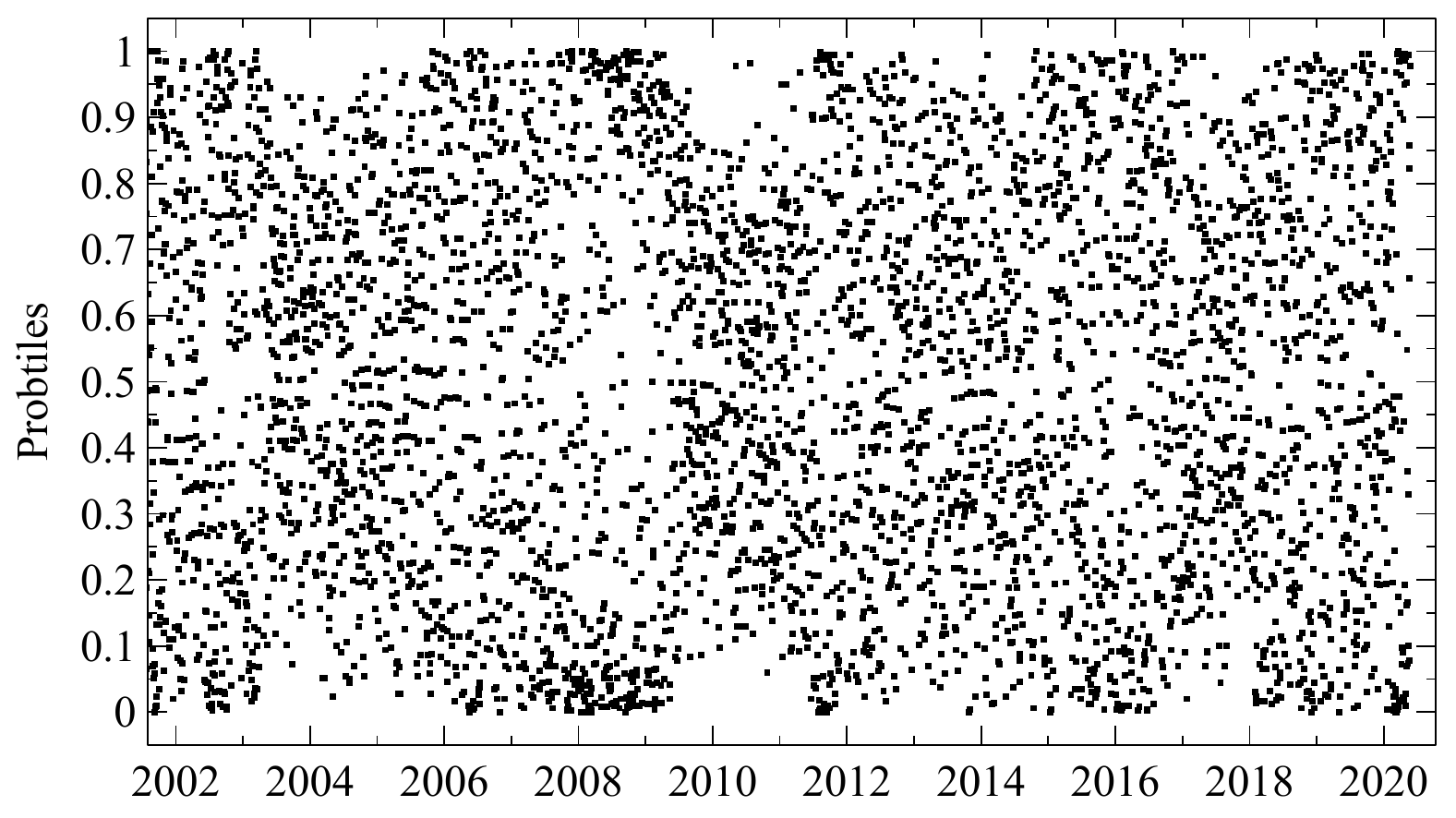}
	
	\caption{The probtiles for the UBS stock, computed using the historical return methodology.
	The non uniformities during some periods are clearly visible, corresponding to a clear over- or under-estimation of the risk.}
	\label{fig:UBS_probtiles_ret}
\end{figure}
\begin{figure}[ht]
	\centering
	\includegraphics[width=1.0\linewidth]{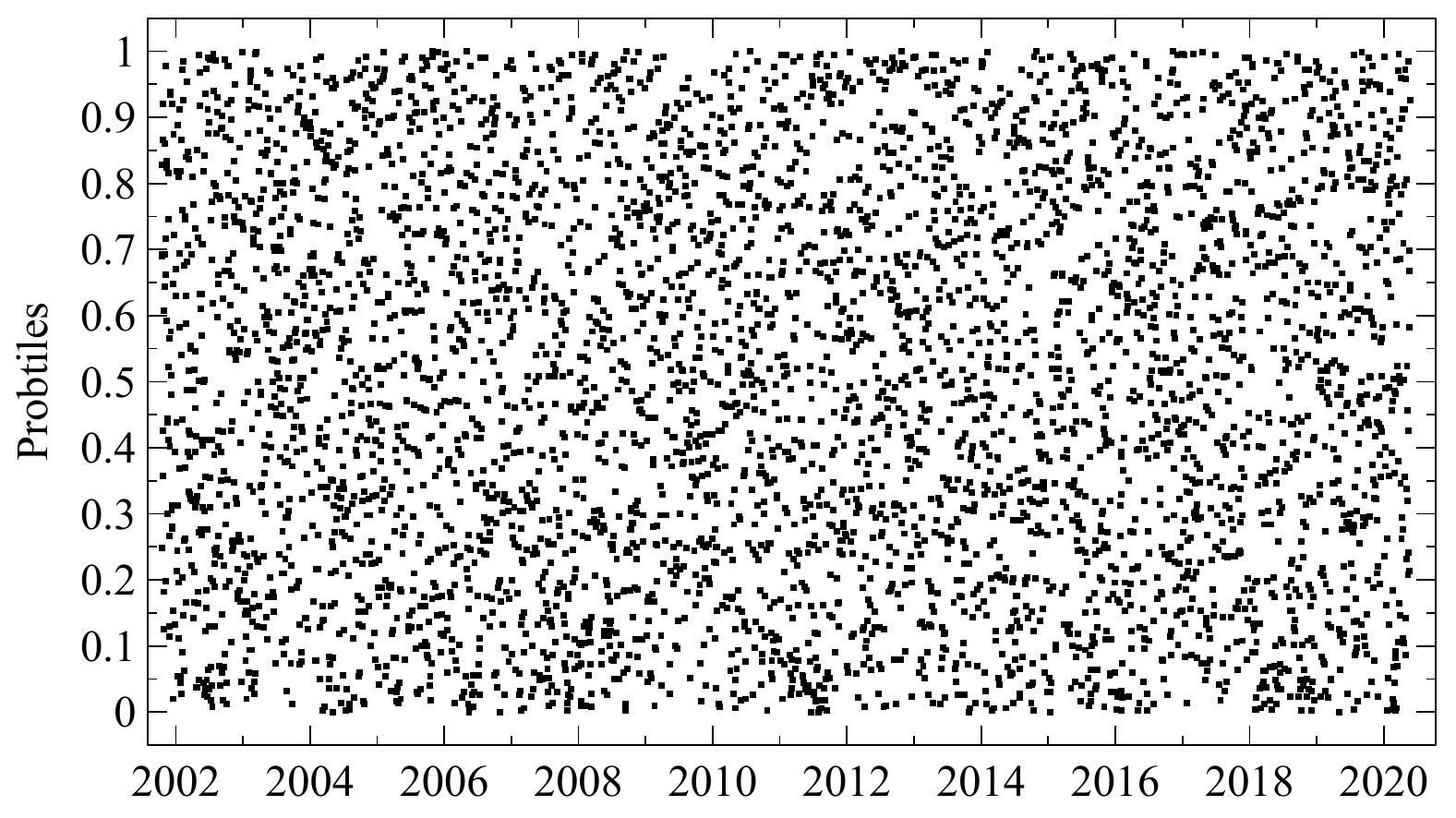}
	
	\caption{The probtiles for the UBS stock, computed using the historical innovation methodology.}
	\label{fig:UBS_probtiles_inno}
\end{figure}
The figure~\ref{fig:UBS_returns} is important since giving the key message in bare form, without any statistics.
These time series are obviously not stationary, and therefore not amenable easily to statistics.
As explained previously, the key idea to obtain a stationary problem is to use a PIT to transform the realised returns into probtiles, using the forecasted distributions.
The figures~\ref{fig:UBS_probtiles_ret} and \ref{fig:UBS_probtiles_inno} show the realized probtiles using respectively the blue and red risk methodologies.
If a risk methodology is correct, the probtiles should be iid with a uniform distribution.
Visually, the points should be uniformly spread in the figure.
This is obviously not the case with the blue methodology, based on historical returns, where clusters are observed.
Even though, the asymptotic distribution, measured on the whole sample, could be uniform.
Hence, testing only the marginal distributions is a fairly weak test of the adequacy of a risk methodology.
Let us emphasize that the blue risk methodology can be off for periods up to years, during which the risk is systematically under- or over-valued.

The core idea for the tile test is to check the number of points on sub-tiles of the whole sample.
The whole sample is divided regularly in $T_t$ intervals along the $t$-axis, and $T_z$ intervals along the probtile $z$-axis, creating $T= T_t \,T_z$ tiles with an equal area. 
In each tiles, $N/T$ points are expected, with $N$ the total number of points in the sample, and $T$ the number of tiles.
The actual number of points on a given tile is $n_i$, and $dn_i = n_i - N/T$ should be a random variable with zero mean.
If the risk model is perfect, the variance of $dn_i$ should be in line with the theoretical model of random points.
If not, the risk model can be rejected.
In the literature, similar tests have been used to test random generators, to check that, in a sequence of draw $x_i$, the vectors $(x_i, x_{i+1}, \cdots, x_{i+n})$ are uniformly spread in a unit cube $[0, 1)^n$ \citep{Knuth.1998}.
This test is of interest in finance since it is sensitive to both the uniformity in $z$ and to the time dynamics.
When applied on Fig.~\ref{fig:UBS_probtiles_ret}, the fluctuations between tiles are larger than with Fig.~\ref{fig:UBS_probtiles_inno}, and the hypothesis of uniformity should be rejected for the first methodology.
A slight complication is that the points are deterministic in time with one point per business day, and random in the $z$ direction.
This is important to derive an analytical benchmark, but for several reasons detailed in this paper, we will only use numerical benchmarks.

This test is used with empirical time series in order to check some major methodologies used to compute financial risk.
A main outcome is that the dynamical part is crucial in delivering a good risk evaluation, then the distributional model is important.
This conclusion is in-line with \citep{Zumbach.backtesting}, where the lagged correlations of the probtiles were tested, yet the tile-test gives a clearly more powerful assessment of the dynamics. 
Interestingly, the bulk of the literature on back-testing focusses on the marginal distributions computed with the full sample, a quantity that is indirectly sensitive to the dynamics through the choice of the sample time boundaries.
This also contradicts the belief of many practitioners who are using long samples of historical returns, inducing a very slow dynamics for the volatility.
The resulting lack of reactivity cost dearly to some banks during the sub-prime crisis in Fall 2008, whereas the volatility started to raise already at the end of 2007, following a long period of decreasing volatility.
Not able to evaluate properly the risks with its dynamics, they kept their positions until September 2008, then made the headlines of newspapers.

The plan of this paper is as follows.
The notation and definition of the tile test is presented in the next section. 
Section~\ref{sec:BenchmarkMonteCarloOnZ} defines the benchmark following the idea that the probtiles $z$ should be iid in $U(0,1)$.
The motivation for the tilings used in this work are presented in Sec.~\ref{sec:tilling}, the risk methodologies in Sec.~\ref{sec:riskMethodologies}, and the empirical data sets in Sec.~\ref{sec:dataSet}.
The problems related to the benchmark are addressed in Sec.~\ref{sec:randomWalkBenchmarks}, first by using the usual benchmark, then by introducing 3 benchmarks appropriate for different risk methodologies.
The properties of these benchmarks are analysed in Sec.~\ref{sec:RandomWalkBenchmarks_Properties}, showing why some test results are too good to be true.
Equipped with the appropriate benchmarks, an empirical study is done for the 1 day and 10 days risk horizons in section~\ref{sec:empiricalResults_1d} and \ref{sec:empiricalResults_10d} respectively.
Finally, the conclusions are presented, on the tile test and on the components required to construct a risk evaluation.

\section{General set-up for the tile test}
\label{sec:generalSetUp}
The notation is as follows.
The risk horizon is $\DT$. 
The total number of points in the sample is $N$.
The number of tiles in the probtile $z$ and time $t$ directions are $T_z$ and $T_t$ respectively, with the total number of tiles $T = T_z \, T_t$.
The number of points in one given tile $i$ is $n_i$.

The time axis is divided regularly from the start to the end of the sample.
The length of the tiles in the $t$-direction is denoted with $\DT_\text{tile} = \DT_\text{sample}/T_t$, with $\DT_\text{sample}$ the sample length in year.
Consider now a column of tiles, spanning some time interval.
The probability for one point to fall in one given tile in the $z$ direction at a given time $t$ is $p = 1/T_z$.
The statistical tests involve the total number of points in the column of tiles.
Due to holidays, the number is slightly changing, and should be counted from the number of valid data points in each column. 
The total number of points in a column of tiles, in the time span for these tiles, is $N_t$ with $N_t \simeq N/T_t$.

The general idea for the tile test is that the points should be uniformly spread in the tiles, and we want to measure the deviation from uniformity.
On a given tile indexed by $i$, the number of points is $n_i$, with the expected mean $\mu_t = N_t/T_z$.
The random variable
\[
\dn_i = n_i - \mu_t
\] 
measures the difference between the actual number of points in the tile $i$ and the mean.
Its sample standard deviation $\tt$ is given by
\[
   \sigma_{\dn} = \sigma_{\dn}(T, \DT) = \sqrt{\frac{1}{T} \sum_i \dn_i^2}
\]
and measures the fluctuations of the point's count between tiles.
The point's count standard deviation $\tt$ is our base random variable, depending on the number of tile $T$, on the time series, on the risk horizon $\DT$, and on the risk methodology used to compute the \textit{ex ante} probability distribution.
Our goal is to construct a statistical test based on $\tt$.
Our null hypothesis is that the probtiles are iid with a uniform distribution, and this induces the null distribution for $\tt$. 
Following the standard statistical testing procedure, a methodology can be rejected if the fluctuations are too large compared to the theoretical distribution obtained with the null hypothesis.

The usual approach is to demonstrate (or assume) an asymptotic distribution with the null hypothesis, then to compute the mean and standard deviation of $\tt$.
Yet, this usual standardization strategy has the following limits.
\begin{itemize}
	\item It is more natural to take $\tt$ as the base random variable to construct a test, but analytical computations can be done only on $\tt^2$. 

	\item For the risk horizon $\DT$ larger than 1 day, the one-day sampling of the probtiles is correlated.
	Unfortunately, the theoretical distribution for $\tt$ or $\tt^2$ is then unknown.
	A possible remedy is to sub-sample the data every $\DT$-points, but this reduces the sample size by the same factor.
	As the back-test gets increasingly difficult for growing $\DT$, we would like to keep as many data points as possible.
	
	\item Stocks with low liquidity have many days with zero return, essentially due to no trading. 
	Many stocks have also quite low prices, making the minimal price increment apparent for small price changes. 
	Both effects are often compounded.
	For such case, the return distributions become singular, with peaks at zero and at the price increments, and the subsequent probtile distributions are not uniform in the neighbourhood of $z = 1/2$.
	Unfortunately, many stocks are in this case, and it is important to validate risk evaluations also in such cases, and not only for indexes or major stocks.

	\item Depending on the algorithm used to compute the risk forecast, a subtle small negative auto-correlation makes the results better than the benchmark. 
	This point is important to interpret the statistical test results, and is discussed extensively in Sec.~\ref{sec:randomWalkBenchmarks}
\end{itemize}
For these reasons, we decided to use exclusively a Monte Carlo approach, simulating $\tt$ as computed from random walks with constant volatility and with normal iid returns, in order to have a reference distribution for the null hypothesis.
Then, the value for $\tt$ is computed for actual data and risk methodology, and the rejection probability is computed numerically with a Probability Integral Transform (PIT) using the null distribution for $\tt$.
For stocks, in order to deal with the slow trading and price granularity, we censor the tiles around $z = 1/2$ which can be easily done in a numerical scheme.
These computations are explained in more details is the following section.

\section{Base Monte Carlo benchmark}
\label{sec:BenchmarkMonteCarloOnZ}
The rejection test using a base Monte Carlo benchmark is build as follows, described for $\DT = 1$.
A sample of random probtiles is drawn from a uniform distribution, one per day along the time axis, and with the same sample length as in the empirical data size $N$.
The tile test statistics $\ttMC$ is computed on this sample, possibly including a censorship around $z=1/2$ if used on the empirical data (i.e. for stocks).
This is done for the range of tile numbers $T$ that we want to use.
The procedure is repeated $\nMC = $ 500 times in order to obtain a numerical estimate of the cumulative distribution $\cdf_\text{MC}(\tt)$ for $\ttMC$.
For an empirical value $\ttEmp$, the tile test statistics is given as
\begin{equation}
	p = 1 - \cdf_\text{MC}(\ttEmp).
\end{equation}
This quantity measures the probability that a value as large as $\ttEmp$ or larger is observed according to the null hypothesis that the probtiles are independent with a uniform distribution.
If $p$ is smaller than a threshold, say 5\%, then the null hypothesis can be rejected.
Essentially, $p$ measures the probability that a risk methodology is correct, and a small $p$ indicates that the risk is incorrectly estimated (or equivalently that $\ttEmp$ is too large compared to a uniform sampling).

This computation is done for a given time series with index $\alpha$, and the corresponding probability $p_\alpha$ is obtained.
In order to have an overall measure for a risk methodology, the values $p_\alpha$ are computed for a set of series, and the simple mean gives the probability that the methodology evaluates correctly the risk for this sample
\begin{equation}
  p_\text{methodology} = \frac{1}{n} \sum_{\alpha} p_\alpha.
\end{equation}

When sampling daily a risk computation at a risk horizon $\DT$ larger than one, the points separated by less than $\DT$ are correlated due to the overlap.
Thus, the Monte Carlo simulation needs to be adapted to include the correlations induced by the overlap.
Yet, another subtle problem occurs with this test, which need an extension to be based on normal random walks.
This is discussed in Sec.~\ref{sec:randomWalkBenchmarks}, and this extension provides naturally for a benchmark at $\DT$ larger than 1 day, while being equivalent to this one for $\DT = 1$.

\section{The choice for the tiling}
\label{sec:tilling}
The choice for the tiles are free, but the overall idea is to get a sequence of coarse to fine tiles, in order to test risk methodologies at various scales.
We have tried a few tilings, in particular increasing simultaneously the division in the $z$ and $t$ directions.

Yet, the most interesting test is in $t$, since depending on the dynamics of the market and of the risk methodologies. 
After several exploratory studies, we singled out the following tiling: a fixed division in the $z$ direction with $T_z = 8$ tiles, and an increasing number of tiles in the $t$ direction. 
The number of tiles in $t$ follows a geometric progression with a factor $\sqrt{2}$.
The number of tiles goes from $T_t = 1$ (i.e. no division, or the full sample in $t$) to a maximal number of tiles corresponding to have no less than 2 points per tiles (so roughly 1 month).
For the figures, the horizontal axis is the time length of the tiles, expressed in year.
This tiling and representation allow to have a diagnostic of a risk methodology as function of the characteristic time length for the test. 

In the figures below for the tile test, the left side of the graph, corresponding to short tiles, is mainly sensitive to the dynamic, while the right side for the graph, corresponding to long tiles, is mainly sensitive to the asymptotic distribution of the probtiles.
Many risk evaluations used in the following empirical study use a 2 year trailing window, which is apparent in the graphs at the centre.

\section{The risk methodologies}
\label{sec:riskMethodologies}
The users typically focus on VaR or ES, but the core object produced by a risk methodology is the (forecast for the) probability distribution for the losses (or for the returns, up to a trivial sign).
A risk methodology is a recipe to construct this forecast, in practice in a multivariate setting, albeit we will investigate only the univariate case.
At the core level, the risk methodologies can be based on the returns or on the innovations.
Since the ones based on the returns are simpler, let us present them first.

The methodologies based on the returns typically ``just'' resample past returns, using a sample of length $N_\text{hist}$.
Over the last $N_\text{hist}$ days, the daily returns are computed, and they are spanning the forecasted distribution.
For a risk horizon $\DT$, the daily returns are scaled by $\sqrt{\DT}$.
A variation is to compute the returns over $\DT$ (using a sample of $N_\text{hist} + \DT$ past prices), and no scaling.
We have investigated these 2 methodologies.
\begin{itemize}
	\item \textbf{Historical returns @ 1d} (with a $\sqrt{\DT}$ scaling)
	\item \textbf{Historical returns @ $\DT$} (without scaling)
\end{itemize}
For the empirical investigation, we have used $N_\text{hist} = 500$ days, corresponding roughly to 2 years.

The algorithm simplicity is very appealing, and is the reason for its wide usage. 
Since it is so simple (\textit{``it just resamples past returns''}), it seems void of hypothesis.
This argument is often used, but is wrong.
The key hypothesis is that the returns have a stationary distribution.
But this is incorrect because of the heteroskedasticity.
The figures given in the introduction display the important dynamics of the volatility and of the return distribution.
To be sure: in finance, the return distributions are time dependent for all time series, hence not stationary.

The related volatility forecast is given by the standard deviation of the return sample.
Accordingly, the variance is an equal weighted sum of squared returns over the last $N_\text{hist}$ days.
A large return entering the sample has a weight $1/N_\text{hist}$, and the weight is constant over the $N_\text{hist}$ subsequent days.
Then, this information is abruptly forgotten when the point leaves the trailing sample, possibly leading to an ``echo'' behaviour with a step down in the volatility. 

The methodologies based on the innovations use a process structure to model the volatility dynamics \citep{Zumbach.RM2006_fullReport}.
At the time $t$, the base equation is
\begin{equation}
    r(t+\dt) = \widetilde{\mu}(t) + \widetilde{\sigma}(t)\,\epsilon(t+\dt)\hspace{3em}\text{with}~\epsilon \sim p(\epsilon)
\end{equation}
where $\widetilde{\mu}(t)$ and $\widetilde{\sigma}(t)$ are the forecast for the mean and standard deviation, computed using the information up to $t$.
This equation can also be viewed as a location/size/shape decomposition of the return distribution, with the location and shape considered as predictable, while the shape is stationary.
In a process, the innovations $\epsilon(t+\dt)$ are assumed to have a fixed distribution $p(\epsilon)$, typically normal or Student (with zero mean and unit variance).
In this setting, the predictable and time dependent parts are in $\widetilde{\mu}(t)$ and $\widetilde{\sigma}(t)$, while the $\epsilon$'s are random and have a time independent distribution $p(\epsilon)$.
Since the expected returns are difficult to predict and small, the default $\widetilde{\mu}(t) = 0$ is often used.
Using historical data and given the returns and the forecasts for the mean and volatility, this equation can be expressed for the innovations
\begin{equation}
\epsilon(t+\dt) = \frac{r(t+\dt) - \widetilde{\mu}(t)}{\widetilde{\sigma}(t)}.
\end{equation}
With the innovations computed from historical data, the stationary assumption for the distribution and its shape can be studied.

Depending on the form for $\widetilde{\mu}$,  $\widetilde{\sigma}$ and $p(\epsilon)$, many methodologies can be constructed.
For all methodologies used in this empirical study, a null mean return is assumed $\widetilde{\mu} = 0$.
The remaining components are the volatility forecast, characterized by the shape of the memory kernel, and the probability distribution for the innovations.
\begin{itemize}
	\item \textbf{Risk Metrics '94}: This is the original proposition based an exponential moving average for the volatility and a normal distribution for the innovations\citep{MinaXiao.2001}.
	
	\item \textbf{LM-ARCH + student}: This is the methodology proposed by RiskMetrics \citep{Zumbach.RM2006_fullReport}, using a long memory ARCH (LM-ARCH) process for the volatility forecast and a Student distribution for the innovations.
	In this model, the volatility forecast is inferred from a daily process using conditional expectations for the squared returns \citep{Zumbach.LongMemory}.
	Since derived from a process, the forecasts are consistent for increasing $\DT$, have a non-trivial term structure, and do not involve new parameters or adjustments.
	The forecast is a weighted sum of past squared returns, with weights that decay smoothly with increasing lags.
	This property captures the progressive decay of the information as it recedes into the past. 
	The original distribution is a Student with 5 degree of freedoms (dof), regardless of $\DT$.
	We report in this study the same model but for dof = 6, which provides slightly better results.
	
	\item \textbf{LM-ARCH + normal}: 
	The same volatility model, with a normal distribution.
	
	\item \textbf{LM-ARCH + historical innovations}: The volatility forecast is based on the same LM-ARCH process, and the distribution for the innovations corresponds to the empirical distribution of the historical innovations using a sample of length $N_\text{hist}$.
	For the empirical investigation, we have used $N_\text{hist} = 500$ days, as for the historical return methodologies.
	This methodology has a non-trivial mean in the returns distribution, corresponding to the mean of the historical innovations multiplied by the volatility forecast.
	As for the historical return methodologies, two variants can be studied, first with innovations at 1 day, second with innovations at the risk horizon $\DT$.
\end{itemize}
More methodologies can be crafted along these lines, but the main issue is to validate, or to invalidate, the various ingredients entering the computations.
Beyond a sound mathematical structure and extensive statistical studies of the stylized facts, back-tests become crucial at this point. 

\section{Data sets}
\label{sec:dataSet}
Two data sets are used for the empirical investigations below.
\begin{itemize}
	\item \textbf{Stock indexes} ~~A set of major stock indexes. Number of time series 10, from 1.1.1993 to 29.4.2020, data length 6614 days. 
	
	\item \textbf{FX} ~~Major foreign exchange rates, all against USD. Number of time series 6, from 1.1.1990 to 4.5.2020, data length 7654 points.

%

\end{itemize}
The results have also been verified on a set of commodities indexes and on a random selection of large and small stocks from the Swiss market.
All the results and conclusions are consistent between the various sets.

\section{Benchmarks based on normal random walks }
\label{sec:randomWalkBenchmarks}
\begin{figure}[ht]
	\centering
	\includegraphics[width=0.48\linewidth]{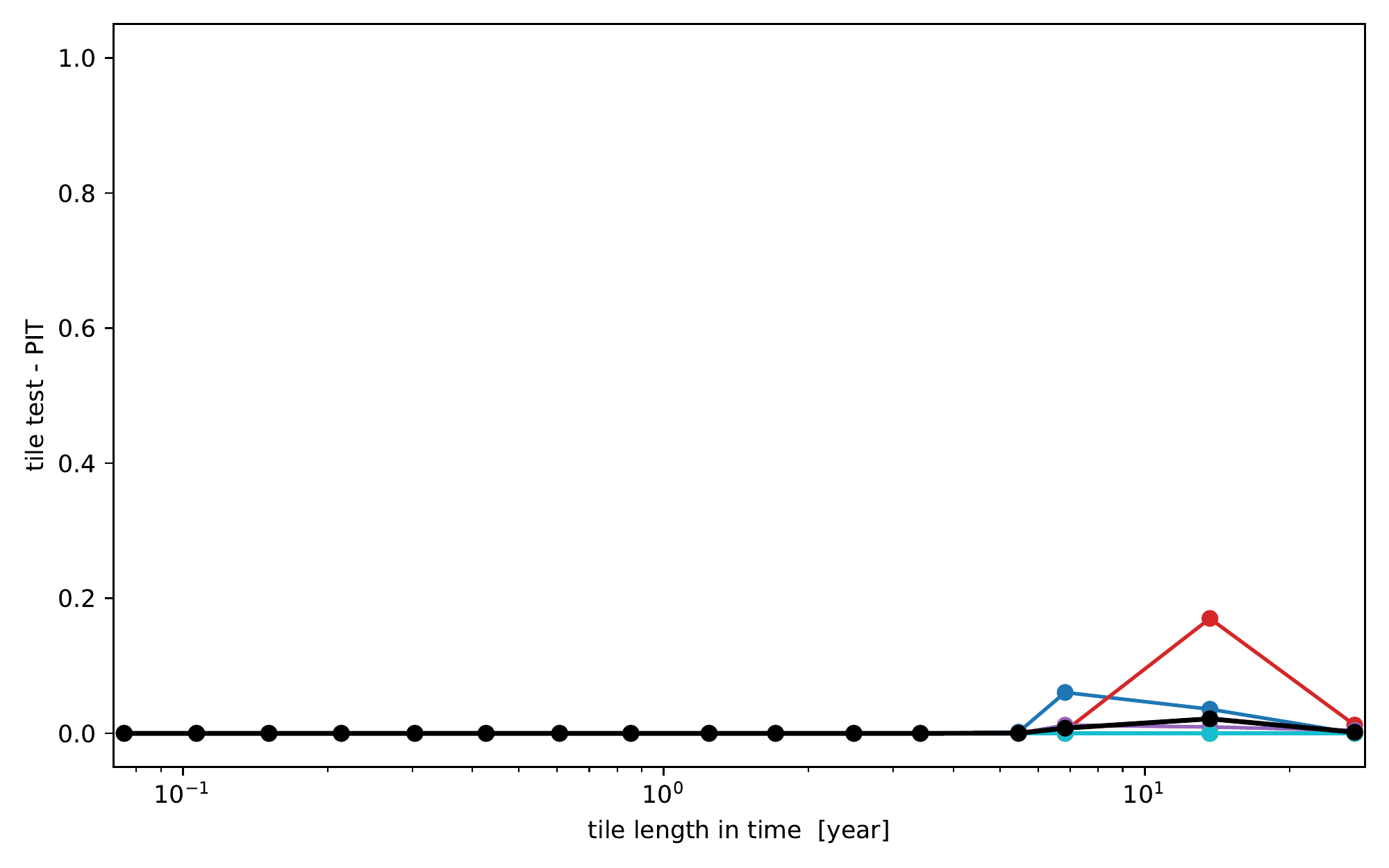}
	\hspace{0.02\linewidth}	\includegraphics[width=0.48\linewidth]{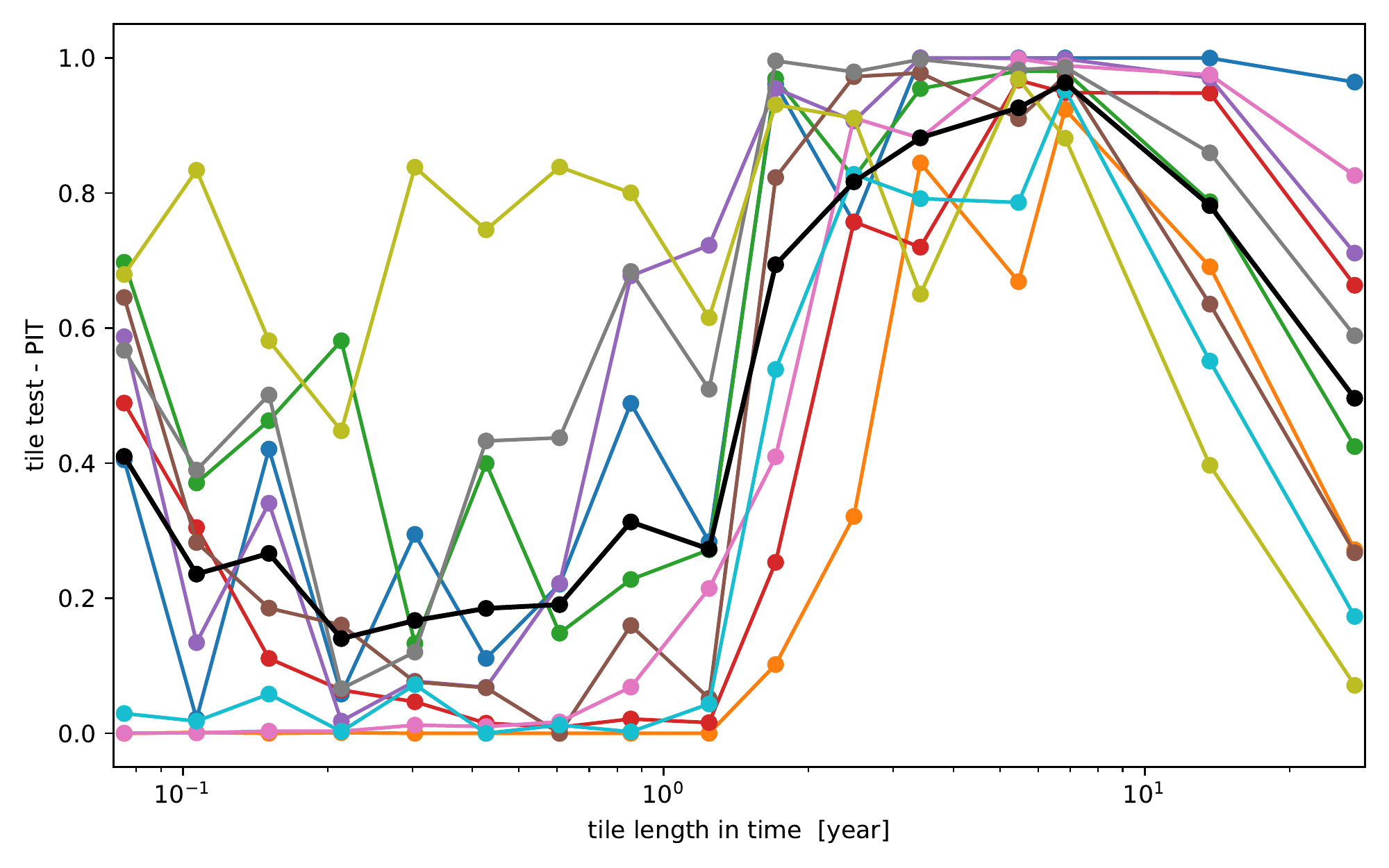}
	\caption{Tile test for the 'indexes' sample at $\DT$ = 1 day, using a uniform benchmark. The left (right) figure is for the historical return (innovation) methodology.}
	\label{fig:tileTestWithUniformBenchmark}
\end{figure}
For the sample of indexes, the figure~\ref{fig:tileTestWithUniformBenchmark} shows the results of a tile test using the iid uniform benchmark. 
The horizontal axis gives the time length of the tile in the $t$ direction, the vertical axis is the probability to observe a value as large or larger than the empirical ones compared to an iid uniform benchmark.
Using the historical return methodology leads to the left figure. 
The values close to zero for the test show that this algorithm can be rejected at almost all time scales.
It is only for very long samples, say more than 5 years that it is not rejected for some indexes.
This shows that asymptotically this methodology could be correct, but a very long sample is required.
Notice that actual risk management operates in the days to months regimes, on the left of the figure, where this methodology is rejected.

Using the LM-ARCH + historical innovation methodology leads to the right figure. 
At first sight, it shows that this algorithm is excellent.
On second sight, the probabilities close to one for long time intervals show that it is always better than a uniform distribution. 
To say it otherwise, the probtiles are distributed systematically more uniformly compared to random uniform points!
Something is wrong, since a perfect model should have values randomly distributed over the $[0, 1]$ interval, and with a mean over many realizations around 1/2.
Here, all empirical time series would reject the theoretical model (using other data sets lead to the same conclusion).
A plot of the probtiles shows nothing special, just random uniform dots.
We checked very carefully our software, to no avail.
But it is clearly not obvious how an actual risk computation can induce probtiles that are more uniform than a theoretical uniform distribution. 

Such a subtle effect is indeed introduced by the trailing sample of historical innovations (and similarly for the historical return algorithm).
In order to understand intuitively the cause, let us assume that one or a few large innovations do occur out of sample. 
They are associated with $z$ values close to one.
These innovations get incorporated into the trailing sample, for the next 500 days in our case.
Subsequently, small and normal innovation values will have $z$ values slightly smaller compared to a theoretical distribution, as long as the large innovations are in the trailing sample. 
The same argument applies for large negative innovations, with slightly larger values for the subsequent probtiles. 
An excess of realized returns around zero produces an excess of probtiles around $z=1/2$ in the trailing sample, then the following returns will be mapped to more extreme values for $z$ (closer to $z=0$ or $z=1$), pushing the distribution toward uniformity.  
This is a type of ``mean reverting'' effect, that produces a slightly different dynamics for the probtiles, while still with a uniform distribution.
The subtle point is that the asymptotic distribution is still a uniform one, but a sequence of such random variables has less variability than independent draws.
Hence, the values returned by the tile test are smaller compared to the ones obtained by iid uniform draws.

The benchmark needs to be adapted in order to include the possible negative lagged correlations induced by an algorithm based on a  ``trailing'' sample. 
This is done by Monte Carlo simulations using a normal random walk with constant volatility, followed by a part that compute the probtiles using a simple algorithm that reproduces the key parts of the actual risk algorithm.
In this form, it is very simple to modify the benchmark to support risk horizons longer than 1 day, while still using the full sample (i.e. not decimating by taking one point every $\DT$ days).
The random sample of returns is computed as follows, for a final sample length of $N$.
A random sample of normal innovations is drawn.
A moving sum of length $\DT$ is performed, with the sum normalized by $1/\sqrt{\DT}$.
The resulting random variables correspond to the (scaled) return at $\DT$, have a normal distribution, a unit variance, a correlations when separated by less than $\DT$, and  a sample length reduced by $\DT - 1$.
On these paths, three benchmarks are computed that reproduce the key parts of the main risk evaluation algorithms.

\begin{itemize}
	\item \textbf{benchmark 1}:  A mapping of the returns with a normal cdf leads to iid uniform random variables with the desired correlation. 
	For this benchmark, the raw Monte Carlo sample length is $\NMod = N + \DT - 1$.
	This benchmark is appropriate when the distribution for the returns or innovations is fixed, for example using a normal distribution.
	
	\item \textbf{benchmark 2}: A trailing sample of 500 daily returns is used to obtain an empirical realisation of the cdf. 
	This cdf is used to map the next out-of-sample return (computed at scale $\DT$ and scaled by $1/\sqrt{\DT}$) to the corresponding probtile $z$. 
	Then, the oldest point in the trailing sample is dropped, the next 1 day return is added in the trailing sample, and the date is moved by 1 day. 
	For this benchmark, the Monte Carlo sample length is  $\NMod = N + \Ntrailing + \DT - 1$ (with $\Ntrailing = 500$).
	This benchmark is appropriate when the distribution for the returns or innovations is based on a trailing sample computed at 1 day.
	
	\item \textbf{benchmark 3}: This benchmark is identical to benchmark 2, but returns at the scale $\DT$ are used in the trailing sample.
	For this benchmark, the Monte Carlo sample length is $\NMod = N + 500 + 2 (\DT -1)$.
	This benchmark is appropriate when the distribution for the returns or innovations is based on a trailing sample of returns or innovations computed at $\DT$ days.
	 
\end{itemize}
Finally, this 1 path algorithm is repeated $\nMC = 500$ times in order to obtain the desired statistics for $\sigma_{\dn}$.
For the present tile test, the tile statistics for the desired tilings are computed on each random path, and the $\nMC$ values are used to build $\cdf_{\MC}(\sigma_{\dn})$ and to compute the corresponding mean $\mu_\MC$ and standard deviation $\sigma_\MC$.
These statistics are obtained for each tilling.
These Monte Carlo cumulative distributions are the final benchmarks for the actual algorithms used with empirical data, and the benchmark 1, 2 or 3 should be chosen according to the algorithm.
We denote by 'adapted benchmark' one of those benchmarks appropriate for a risk methodology.

A few points should be emphasized.
First, at a risk horizon of 1 day, the benchmark 1 is drawing iid normal variables followed by a mapping with the normal cdf, hence uniform iid variables are obtained.
Therefore, the benchmark 1 at 1 day corresponds to the usual uniform benchmark, namely to iid uniform probtiles.
For longer risk horizon, it is similar but with correlations for values separated by less than $\DT$.
For a 1 day risk horizon, the benchmark 2 and 3 are identical,but they are different for $\DT > 1$. 

Second, because the volatility of the random walk is constant and the distribution is fixed to a normal, there is no heteroskedasticity and no fat-tails. 
It is clearly much simpler to forecast risk in this theoretical Bachelier world, whereas real life includes heteroskedasticity and fat-tails. 
Consequently, the benchmarks are tough for actual algorithms operating on empirical time series.
Also, because the volatility is constant, there is no difference in the benchmarks between a methodology based on the returns or on the innovations.  

Third, other statistics can be computed on the Monte Carlo paths, say for example lagged correlations or Kolmogorov-Smirnoff statistics.
As for the tile test, the values returned by the 3 benchmarks can be noticeably different.

Fourth, more complex algorithms can be designed for risk evaluation, say for example a parametric distribution is estimated on the moving sample of 500 returns. 
Such algorithms induce also a dependency on a trailing sample, and the benchmark should be computed accordingly.

\section{Random walk benchmarks: properties}
\label{sec:RandomWalkBenchmarks_Properties}

Using the base set-up, the probability distribution for $\tt$ is investigated.
The sample length is $N = 5052$ business days, corresponding approximately to 20 years of data.
The number of tiles in the $z$ direction is constant $T_z = 8$, in the $t$ direction is ranging from 1 to 256 (at which there is in average 2.5 points per tiles).
The number of tiles increases with a geometric progression in the $t$ direction with reason $\sqrt{2}$.
The results are better presented using a folded-cdf, namely the cdf for cdf $< 0.5$ and 1-cdf for cdf $> 0.5$.
This representation gives a truthful representation for an empirical or Monte Carlo sampling, without using a smoothing kernel.
Furthermore, a logarithmic vertical scale can be used, focusing on both tails of the cdf.

\begin{figure}
	\centering
	\includegraphics[width=0.7\linewidth]{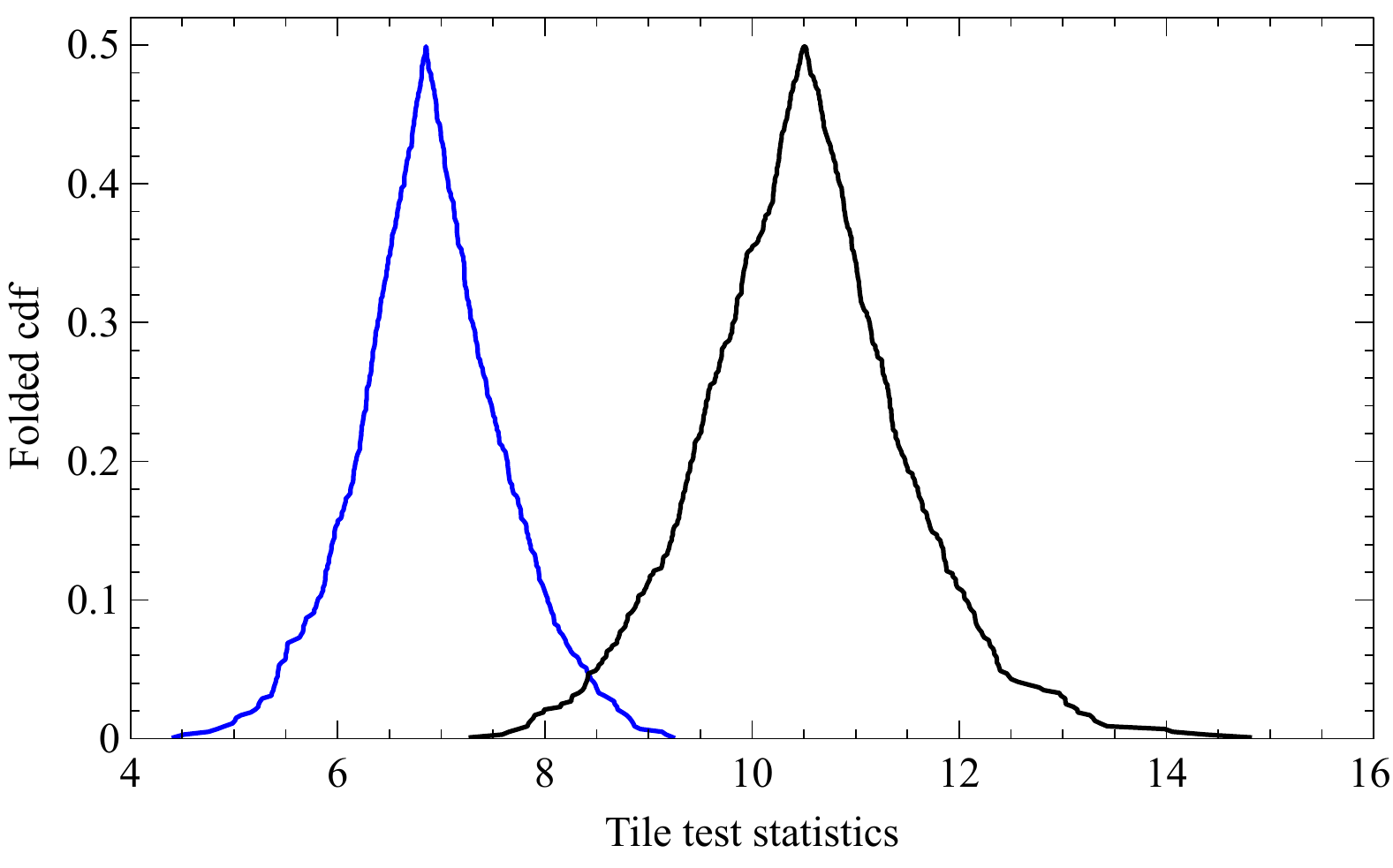}
	\caption{The folded $\cdf(\tt)$ for the tile test $\tt$ for benchmarks 1 (black curve) and 2 (blue curve) for $\DT =1$, with 8 divisions in the $z$ direction and $4$ in the $t$ direction (corresponding to a tile length of 5 years).}
	\label{fig:cdfForBacktest_32}
\end{figure}
The figure~\ref{fig:cdfForBacktest_32} gives the folded cdf for benchmark 1 and 2 fot $\DT = 1$, with a low number of divisions in the $t$ direction.
Both distributions have an almost disjoint supports, with the benchmark 2 displaying markedly lower values for $\tt$, namely fewer fluctuations than the usual iid uniform benchmark corresponding to benchmark 1.
Depending on the tilling, the displacement of the distributions shows the very strong effect induced by the trailing sample for the empirical return distribution.
This difference is the cause for the too good empirical results reported in Fig.~\ref{fig:tileTestWithUniformBenchmark}.

\begin{figure}
	\centering
	\includegraphics[width=0.7\linewidth]{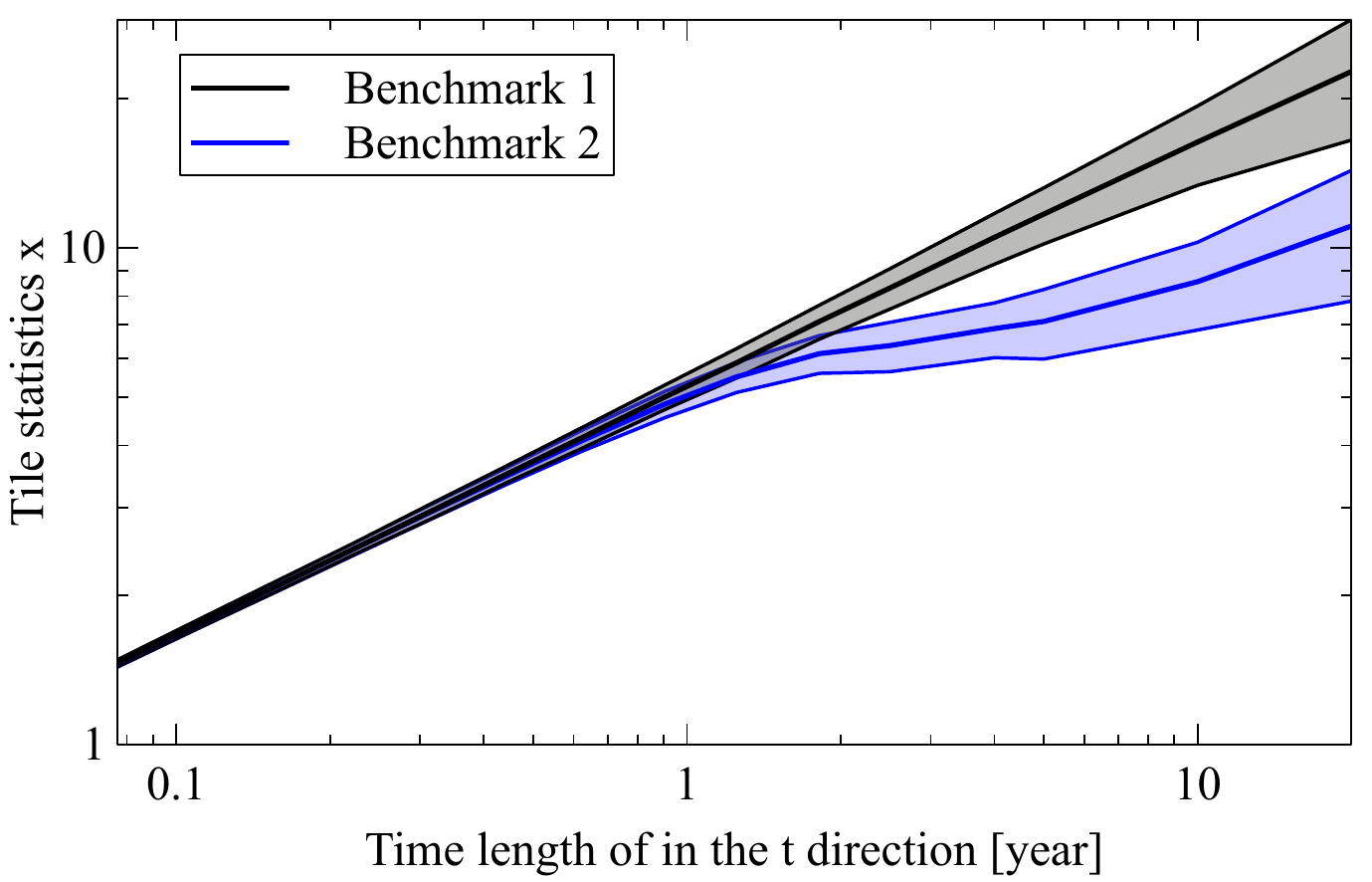}
	\caption{The ``one $\sigma$ range'' for benchmark 1 and 2 at $\DT = 1$, as function of the tilelength in the $t$ direction.
	The coloured areas correspond to a range of $\pm\sigma$ around the mean.}
	\label{fig:oneSigmaRange}
\end{figure}
The core of the distributions is plotted in Fig.~\ref{fig:oneSigmaRange} as function of the tile length in the $t$ direction. 
For increasing tile lengths, the differences between both benchmarks increase, showing the importance of selecting an appropriate benchmark.
The difference becomes large when the tile length is of the order of the trailing sample, in our case 500 business days equivalent to 2 years.
This difference explains the figure~\ref{fig:tileTestWithUniformBenchmark} where the historical innovation algorithm is better than the benchmark for tile length longer than 2 years.

\begin{figure}
	\centering
	\includegraphics[width=0.7\linewidth]{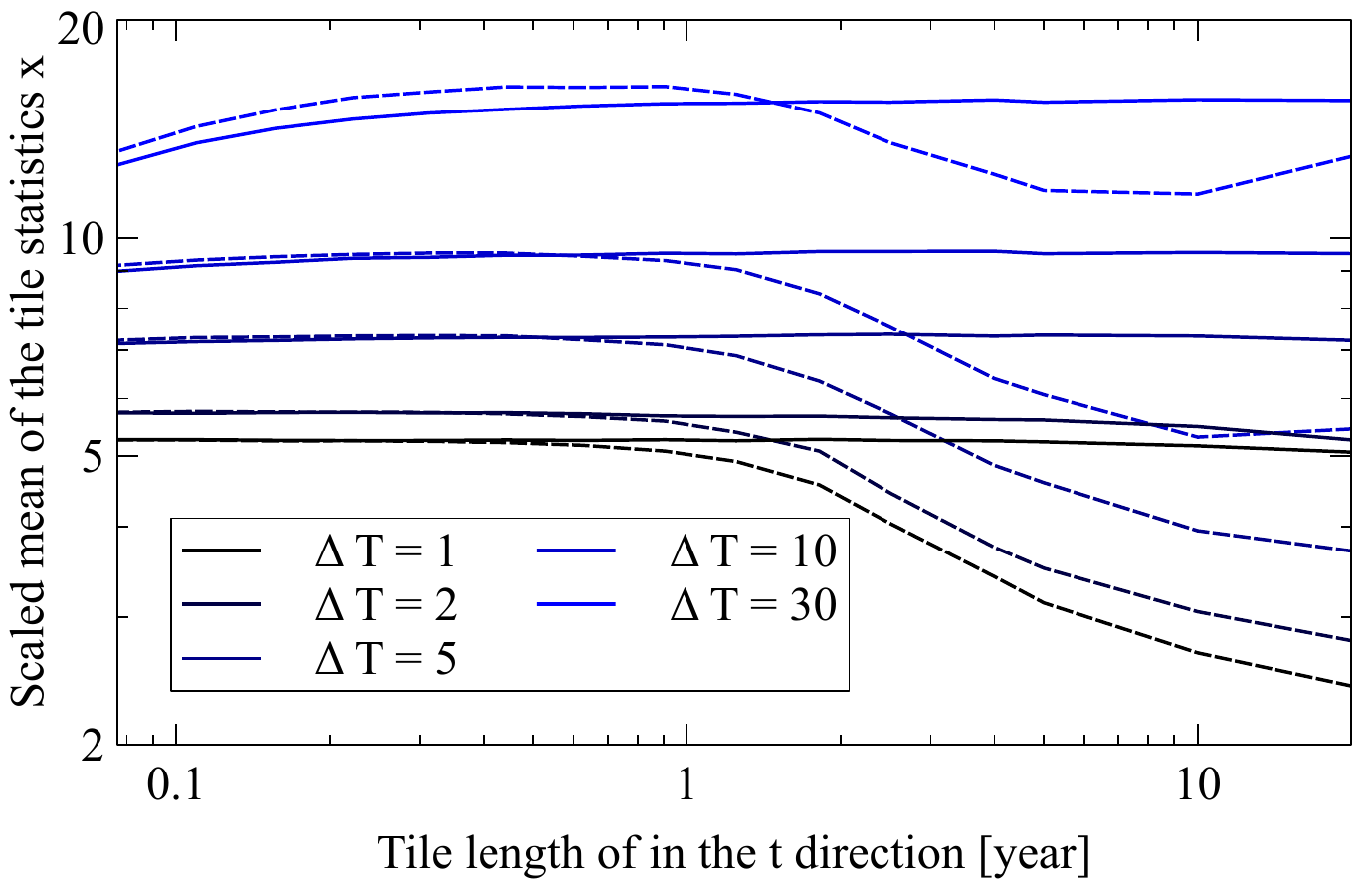}
	\caption{The scaled mean of the tile statistics $\tt$ obtained from MonteCarlo simulations, for benchmark 1 and 3, as function of the tile length in the $t$ direction, for increasing risk horizon $\DT$.
	The scaled mean is $\mu_\MC/\sqrt{\DT_\text{tile}}$, with $\mu_\MC$ the mean of the tile statistics in the Monte Carlo simulations and $\DT_\text{tile}$ the tile length in year.
	The risk horizons are 1 (black), 2, 5, 10, 30 (blue), the benchmark 1 are drawn with full lines, the benchmark 3 with dashed lines.}
	\label{fig:scaling_mean_bench1_bench3}
\end{figure}
Figure~\ref{fig:scaling_mean_bench1_bench3} displays the scaled mean of the tile statistics $\tt$ for increasing risk horizons, for the benchmarks 1 and 3.
This graph shows clearly the different behaviours when the tile length is shorter or longer than the memory length of 500 days $\simeq$ 2 years used in the benchmark 3.
This effect is persistent for increasing risk horizon $\DT$. 
The reduction of the mean for large $\DT_\text{tile}$ originates in the mean reversion induced by the trailing windows, and it occurs regardless of the risk horizon.

The curves are moving upward with increasing $\DT$.
This is due to the reduction  of the effective sample length induced by the correlations of the overlapping returns.
As a first estimate, the effective sample length goes as $1/\sqrt{\DT}$, hence the mean of the fluctuations increases as $\sqrt{\DT}$. 
The actual increase is smaller, of the order of $\DT^{0.25}$ to $\DT^{0.3}$.
This smaller increase is the gain due to the overlapping sample, namely to sample every day even when the risk horizon is longer.

The standard deviation shows a very similar behaviour when scaled as $\sigma_\MC/\DT_\text{tile}$, both in term of the reduction of the variance for the benchmark 3 with increasing tile length, and for the slower increase in $\DT$ than a square root.
These features, for the mean and the standard deviation, lead to a more powerful test, including for risk horizon larger than 1 day.
Indeed, most applications of risk evaluation are for medium to long risk horizons, say 5 days to a few months.
This domain is increasingly difficult to test, and most of the numerical tests are inconclusive. 
The above gains are therefore instrumental at extending risk validation above 1 day.

This section analyses the benchmarks for the tile test statistics, and its dependency on the risk algorithm.
But a similar problem occurs for all statistics applied on a back-test: depending on the risk methodology, the simple benchmark can be (strongly) biased toward accepting a methodology.
Hence, for all statistics, a similar approach should be used, namely to compute a benchmark from the statistics evaluated on random paths with constant volatility and normal returns.


\section{Empirical results for $\DT = 1$ day}
\label{sec:empiricalResults_1d}
\begin{figure}
	\centering
	\includegraphics[width=0.48\linewidth]{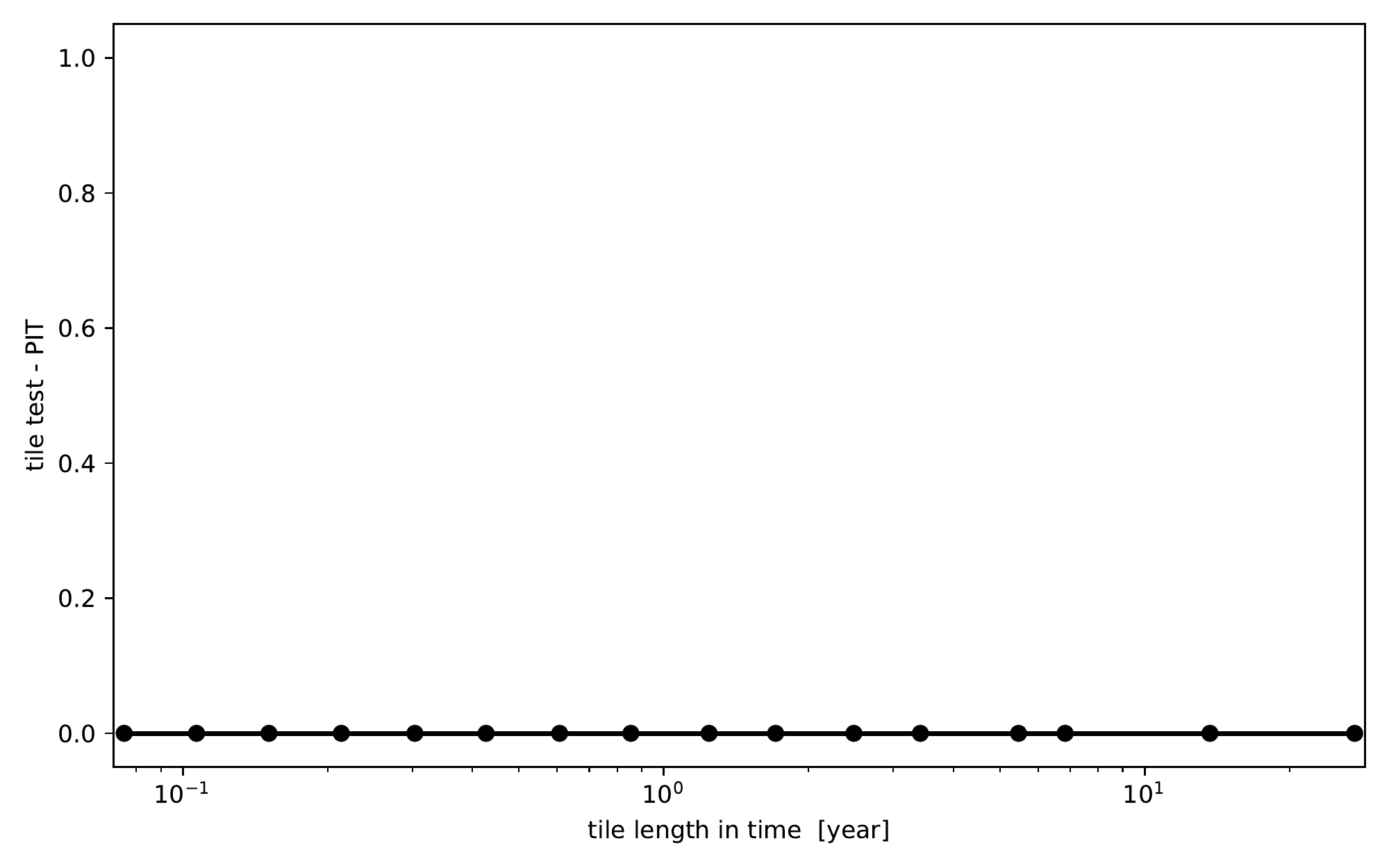}
	\hspace{0.02\linewidth}	\includegraphics[width=0.48\linewidth]{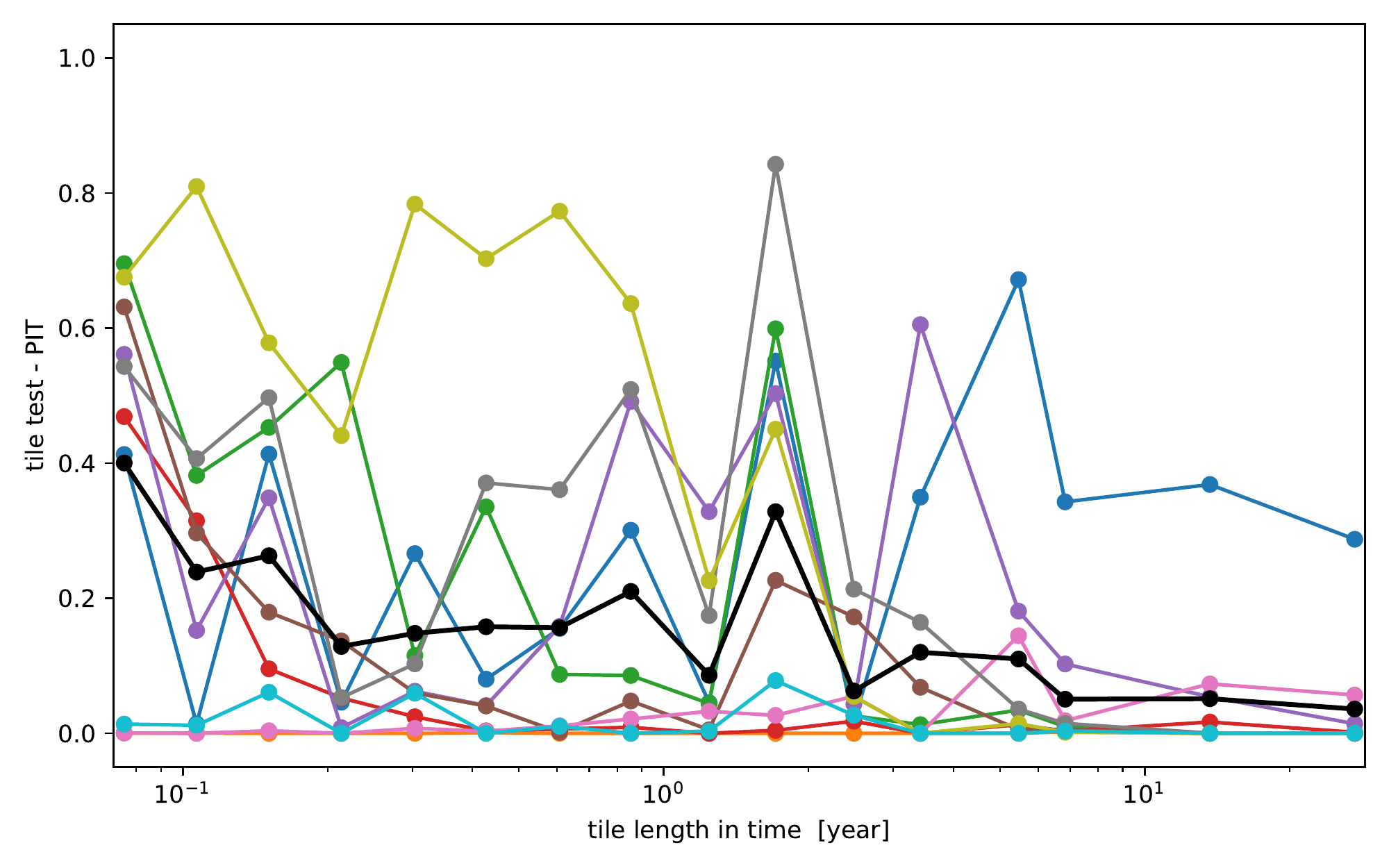}
	\caption{Tile test for the 'indexes' sample at $\DT$ = 1 day, using an adapted benchmark. The left (right) figure is for the historical return (innovation) methodology.}
	\label{fig:tileTestWithBenchmark2}
\end{figure}
The Fig.~\ref{fig:tileTestWithBenchmark2} is identical to Fig.~\ref{fig:tileTestWithUniformBenchmark}, but with the benchmark 3, namely taking into account that both algorithms use a trailing sample of returns or innovations.
Against this benchmark, the historical return methodology is rejected strongly, at all time lengths, and for all series in this sample.
The rejection of this model demonstrates the importance of the dynamics.
Even with a tile length using the full sample (more than  25 years), this algorithm is strongly rejected. 
This means that 2 decades is not long enough to reach an asymptotic regime where the return distributions converge to their long term limits.
By contrast, the historical innovations cannot be rejected for most time series.
It should be emphasized that the test is comparing a normal random walk with constant volatility against real data with heteroskedasticity and fat-tails.
The brutal difference between both methodologies shows that it is important to understand the stylized facts of the financial markets, to have good mathematical models for the time evolution, and only then sound risk evaluation algorithms can be designed.

\setlength{\figwidth}{0.40\linewidth}

\begin{figure}
\hspace{0.4\figwidth} \hspace{0.1\figwidth} Benchmark 1 \hfill\hspace{0.3\figwidth} Adapted benchmark \hfill \mbox{}\\[1ex]
\makebox{\parbox[b]{0.4\figwidth}{\raggedright Historical return\vspace*{12ex}}}
\includegraphics[width=\figwidth]{figures/historicalReturns_at1d/DT_1/tileTest_PIT_bench1_indexes}
\hspace{0.02\linewidth}
\includegraphics[width=\figwidth]{figures/historicalReturns_at1d/DT_1/tileTest_PIT_bench2_indexes}
\\
\makebox{\parbox[b]{0.4\figwidth}{\raggedright  RiskMetrics\vspace*{12ex}}}
\includegraphics[width=\figwidth]{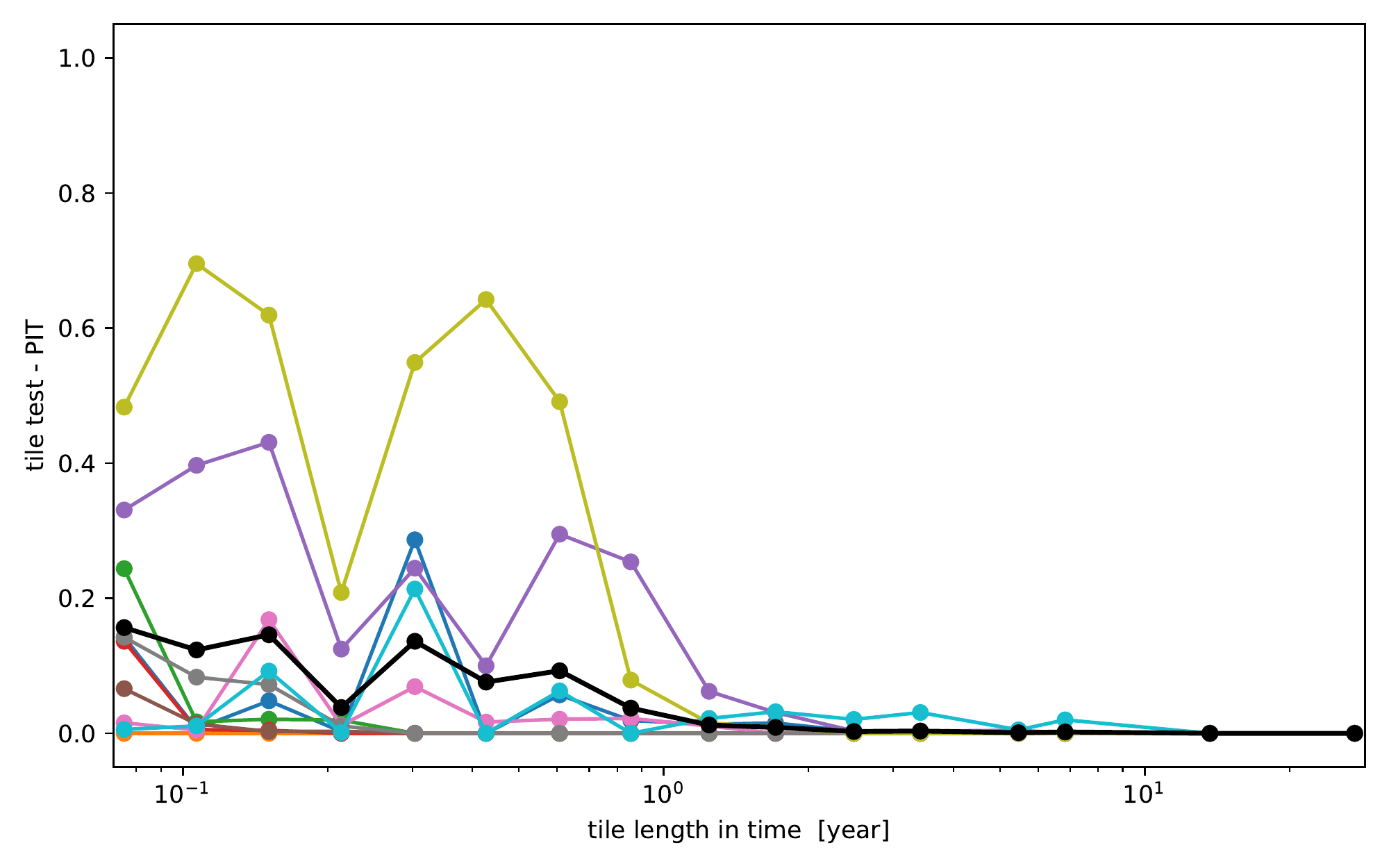}
\hspace{0.02\linewidth}
\includegraphics[width=\figwidth]{figures/RiskMetrics/DT_1/tileTest_PIT_bench1_indexes}
\\
\makebox{\parbox[b]{0.4\figwidth}{\raggedright  LM-ARCH + Student 6\vspace*{12ex}}}
\includegraphics[width=\figwidth]{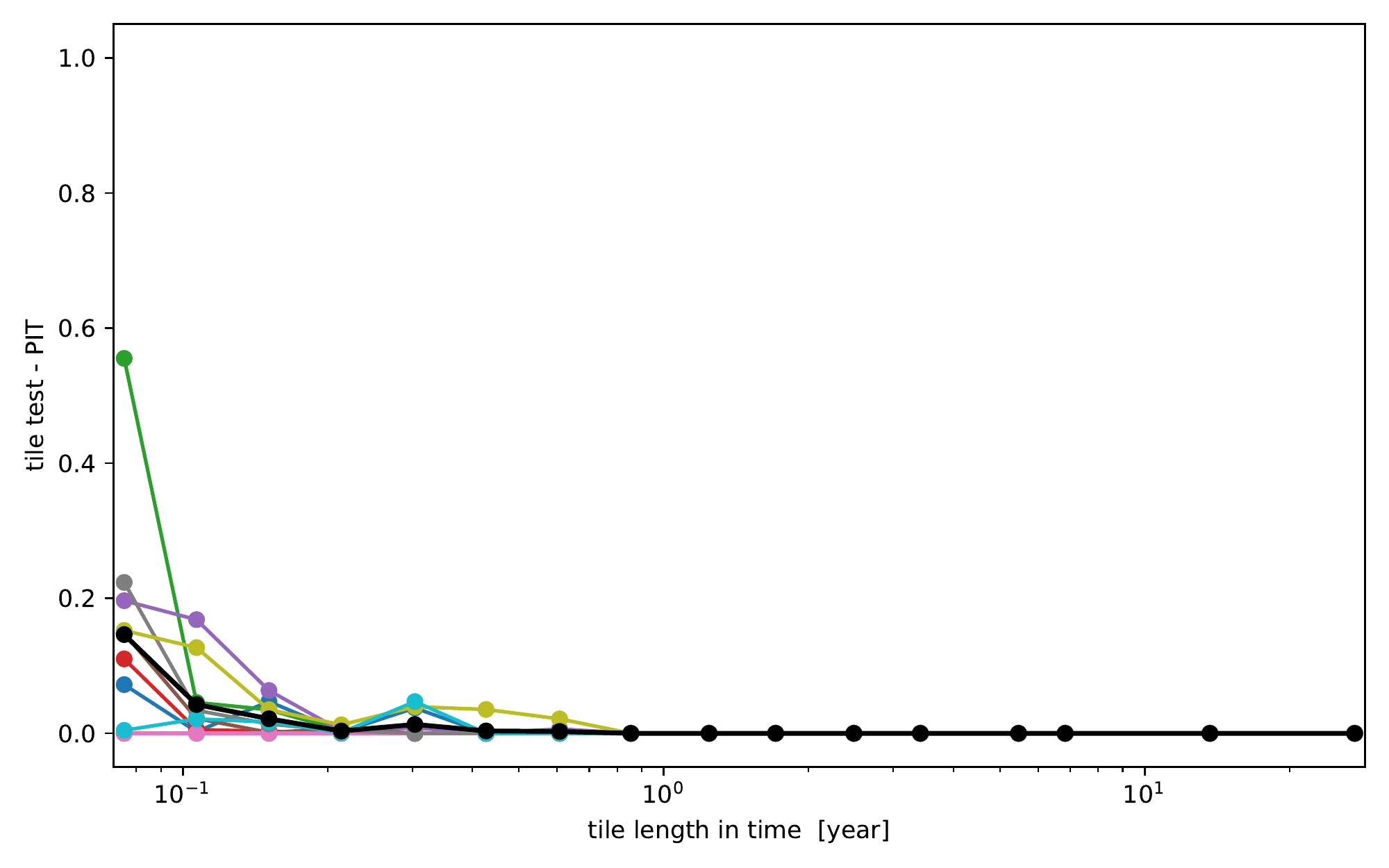}
\hspace{0.02\linewidth}
\includegraphics[width=\figwidth]{figures/LMARCH_Student6/DT_1/tileTest_PIT_bench1_indexes}
\\
\makebox{\parbox[b]{0.4\figwidth}{\raggedright  LM-ARCH + emp.cdf\vspace*{12ex}}}
\includegraphics[width=\figwidth]{figures/LMARCH_empCdf_at1d/DT_1/tileTest_PIT_bench1_indexes}
\hspace{0.02\linewidth}
\includegraphics[width=\figwidth]{figures/LMARCH_empCdf_at1d/DT_1/tileTest_PIT_bench2_indexes}
	
	\caption{Tile test for several risk methodologies, for the 'indexes' sample, at $\DT$ = 1 day, using the benchmark 1 (left column), and the adapted benchmark (right column). The left column allows comparing directly the methodologies using the same benchmark, while the right column allows to compare a methodology against its adapted random walk benchmark.}
	\label{fig:tileTestAtDt1_indexes}
\end{figure}

\begin{figure}
	\hspace{0.4\figwidth} \hspace{0.1\figwidth} Benchmark 1 \hfill\hspace{0.3\figwidth} Adapted benchmark \hfill \mbox{}\\[1ex]
	\makebox{\parbox[b]{0.4\figwidth}{\raggedright Historical return\vspace*{12ex}}}
	\includegraphics[width=\figwidth]{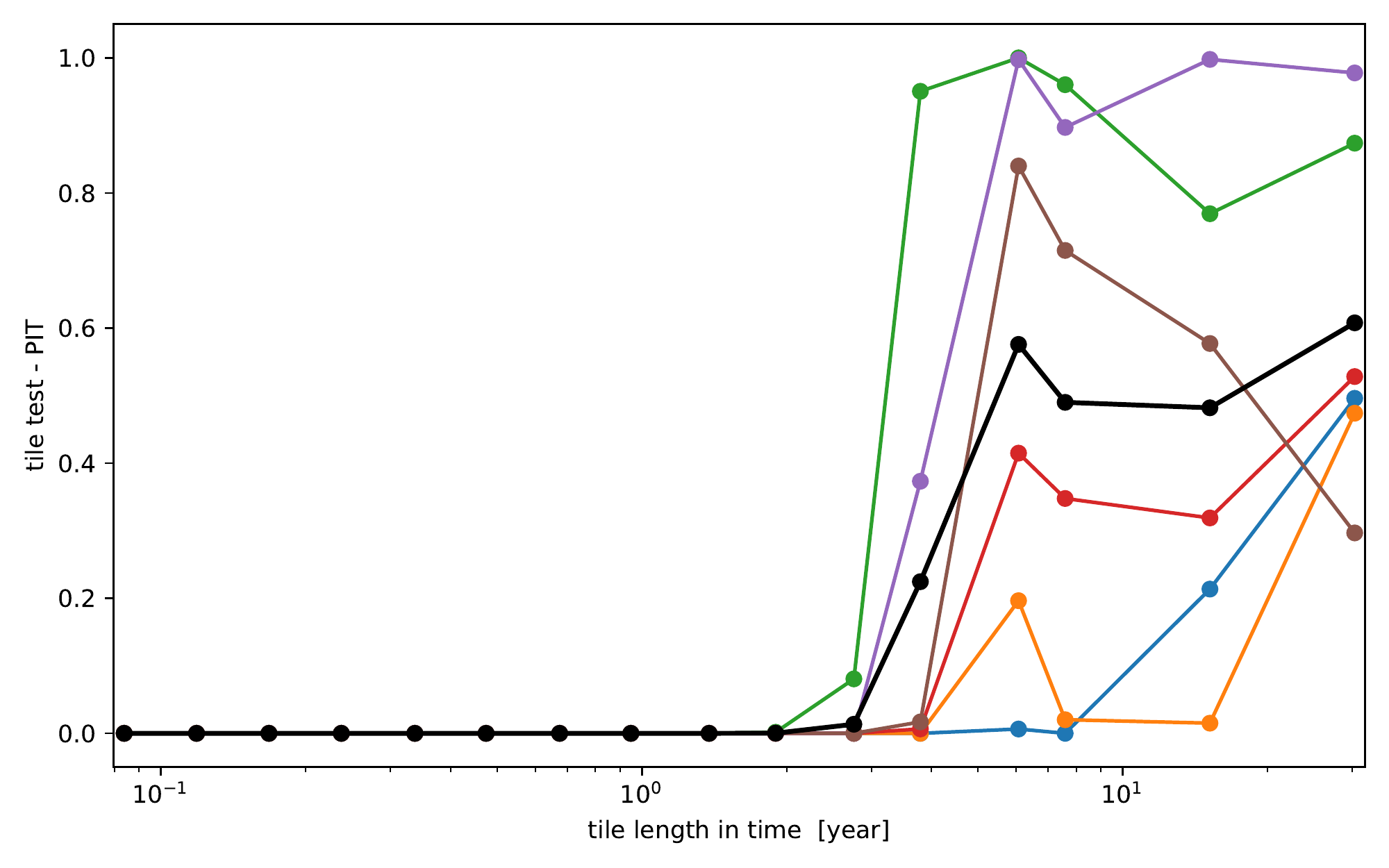}
	\hspace{0.02\linewidth}
	\includegraphics[width=\figwidth]{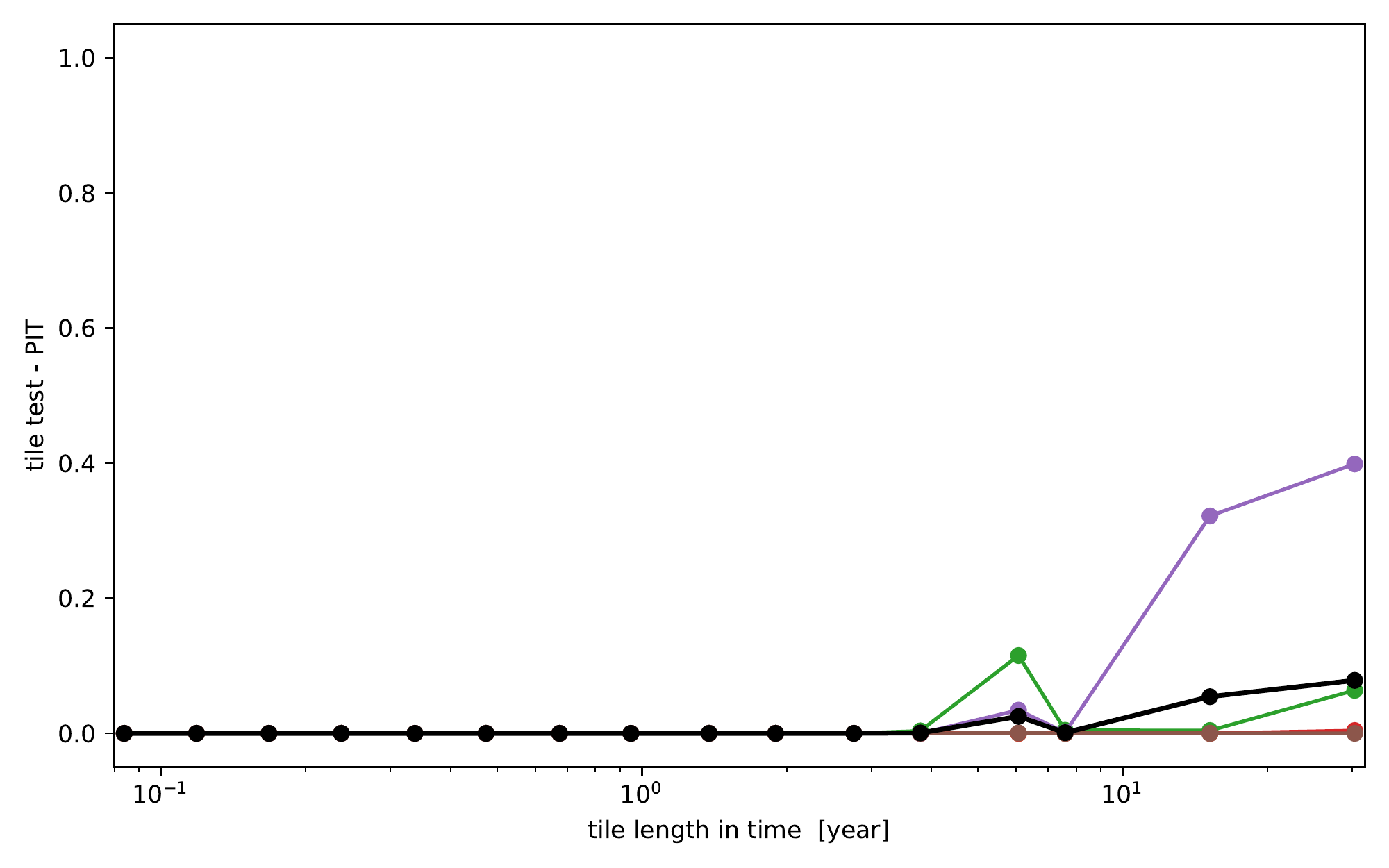}
	\\
	\makebox{\parbox[b]{0.4\figwidth}{\raggedright  RiskMetrics\vspace*{12ex}}}
	\includegraphics[width=\figwidth]{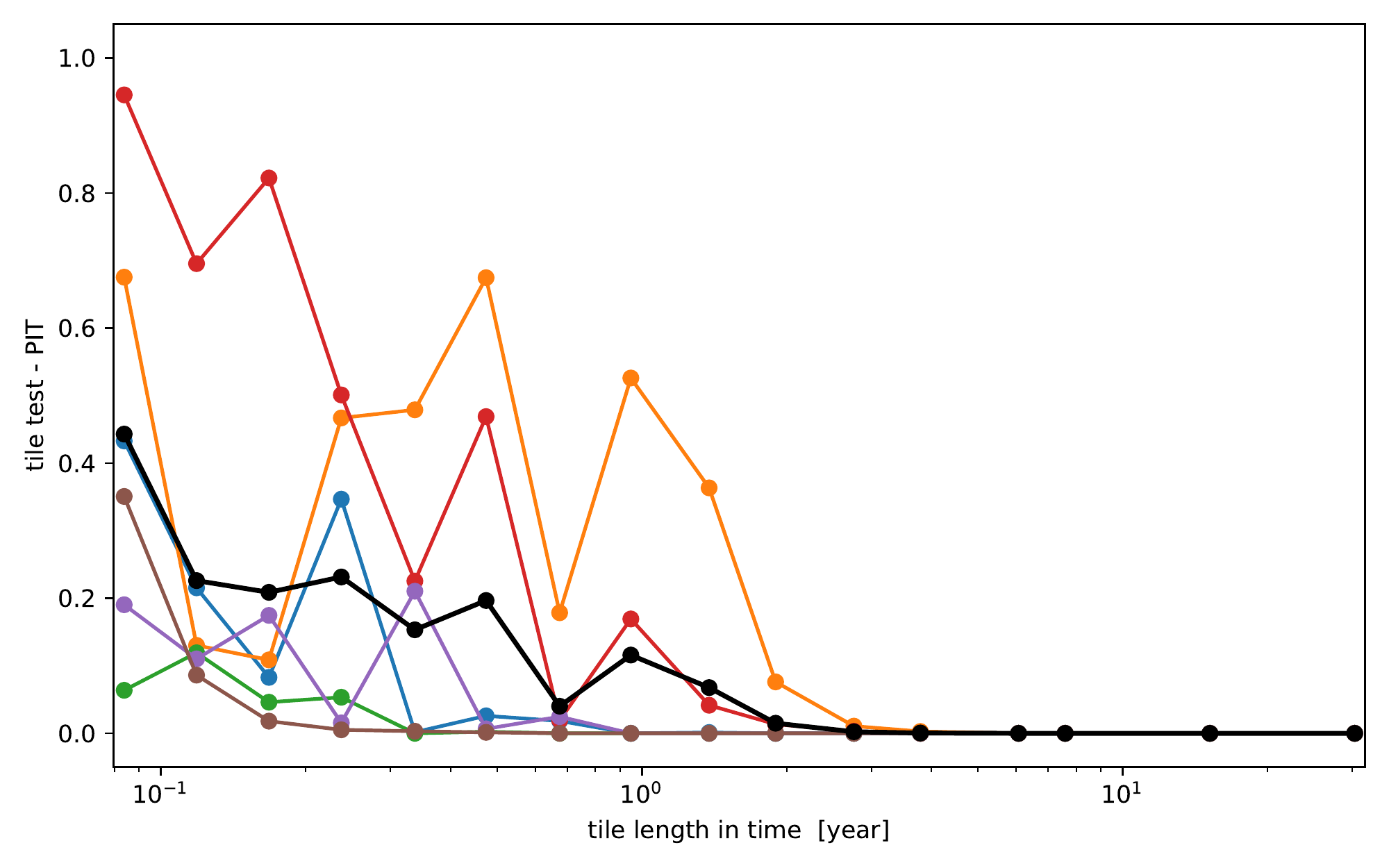}
	\hspace{0.02\linewidth}
	\includegraphics[width=\figwidth]{figures/RiskMetrics/DT_1/tileTest_PIT_bench1_FX}
	\\
	\makebox{\parbox[b]{0.4\figwidth}{\raggedright  LM-ARCH + Student 6\vspace*{12ex}}}
	\includegraphics[width=\figwidth]{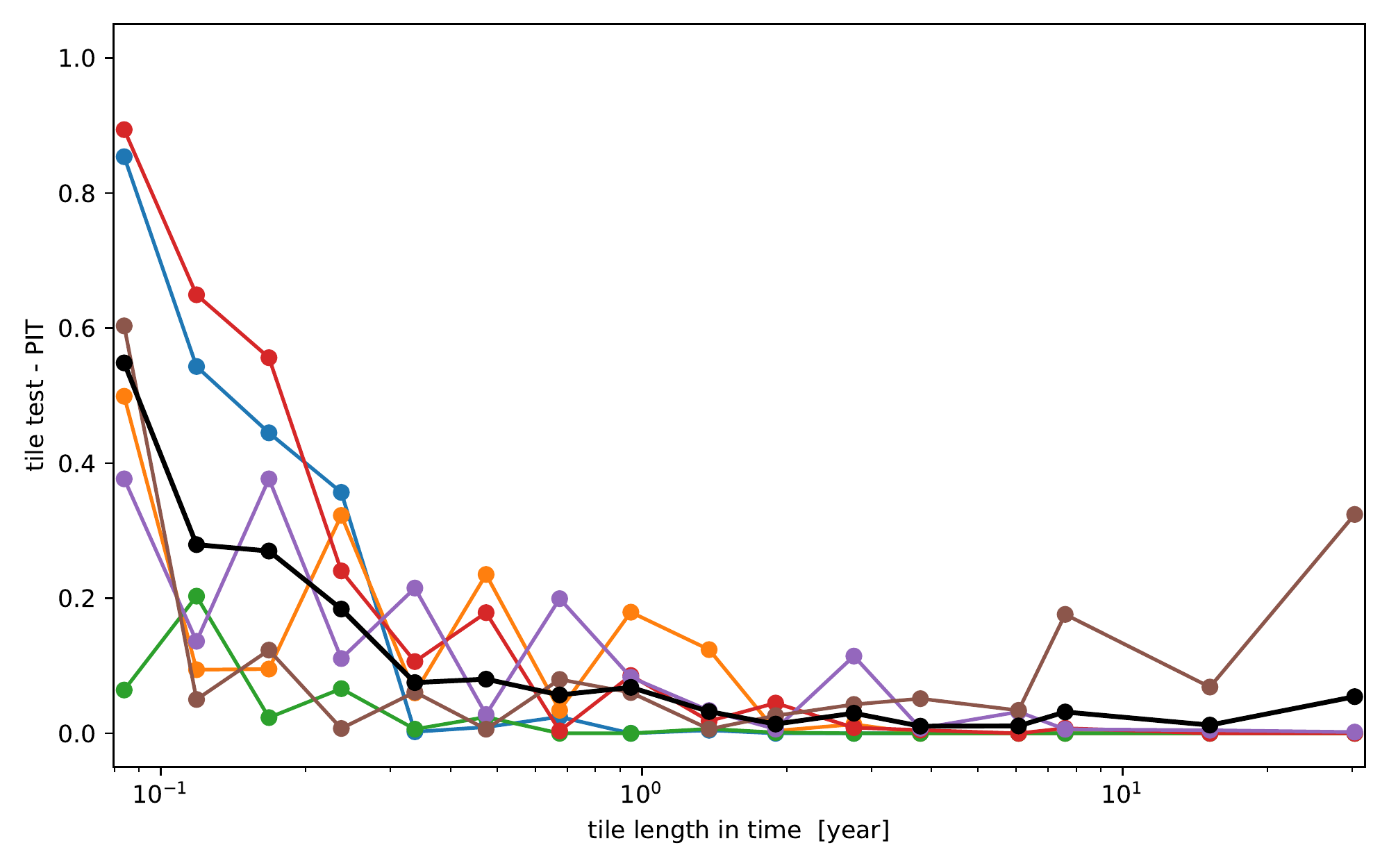}
	\hspace{0.02\linewidth}
	\includegraphics[width=\figwidth]{figures/LMARCH_Student6/DT_1/tileTest_PIT_bench1_FX}
	\\
	\makebox{\parbox[b]{0.4\figwidth}{\raggedright  LM-ARCH + emp.cdf\vspace*{12ex}}}
	\includegraphics[width=\figwidth]{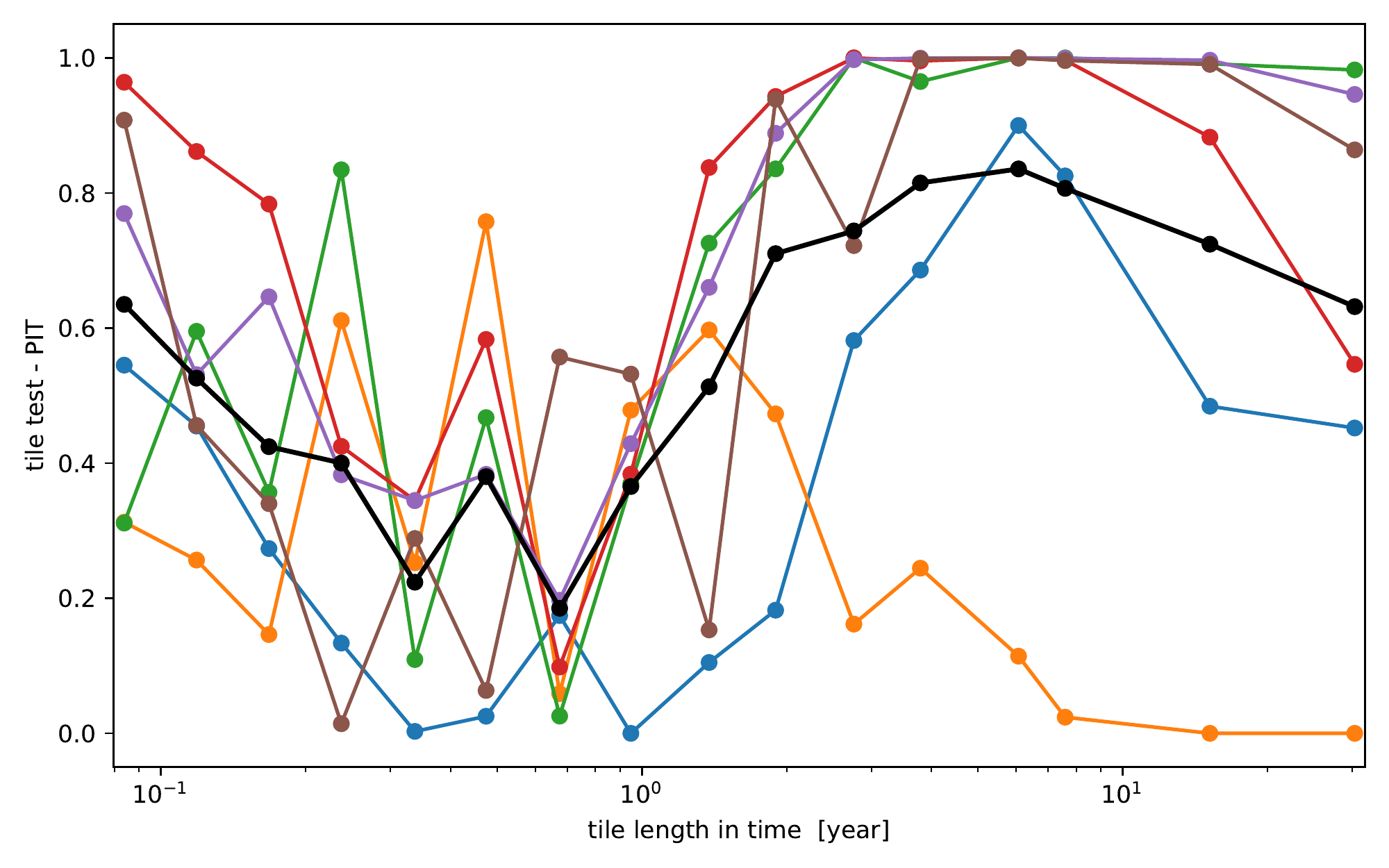}
	\hspace{0.02\linewidth}
	\includegraphics[width=\figwidth]{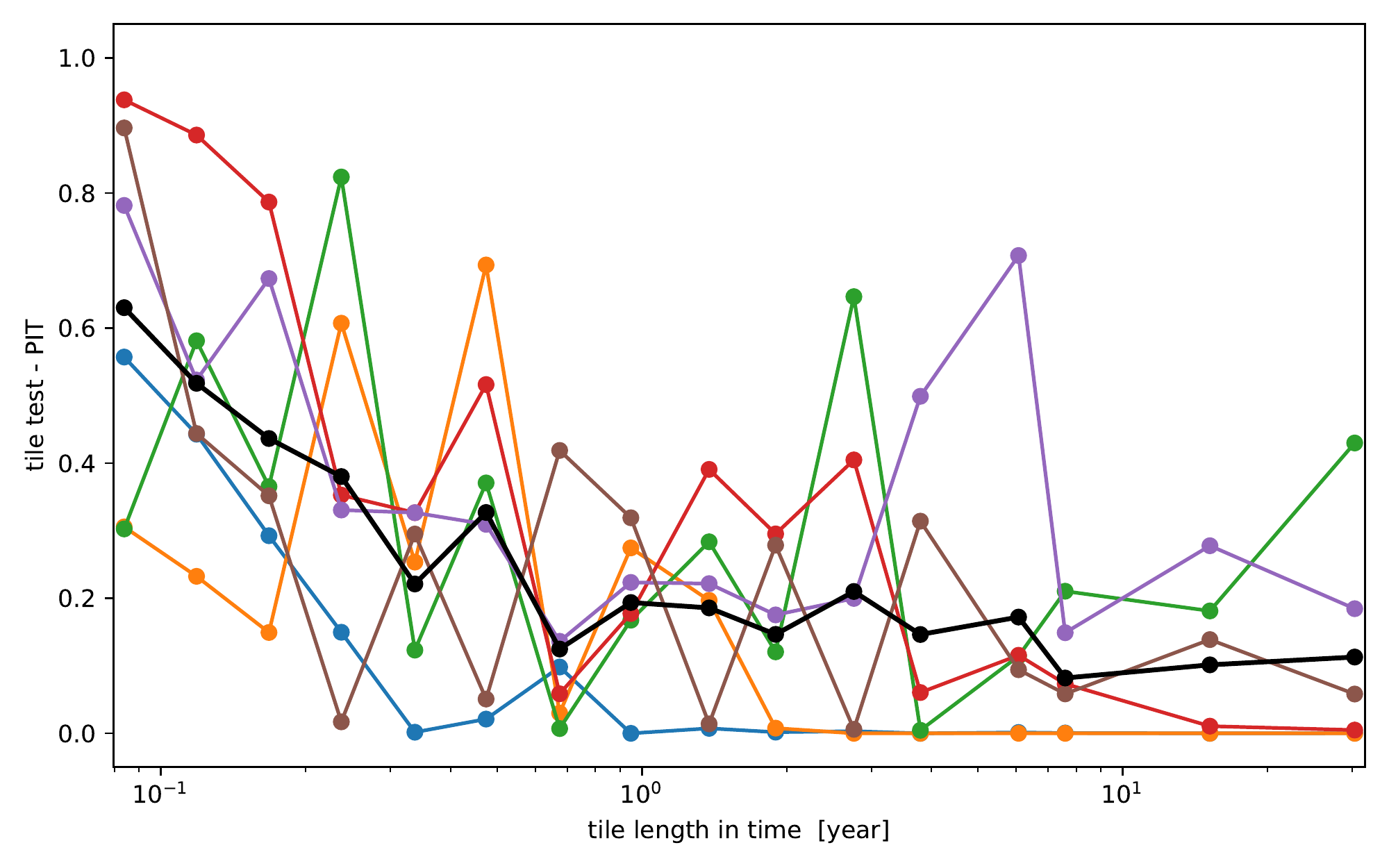}
	
	\caption{Tile test for several risk methodologies, for the 'FX' sample, at $\DT$ = 1 day. }
	\label{fig:tileTestAtDt1_FX}
\end{figure}
Now that we have a good understanding of the interactions between the risk methodologies and the benchmarks, a performance analysis of the main methodologies can be made using the tile test. 
The main ingredients we want to explore are returns or innovations based methodologies, the model for the volatility forecast, and the probability distribution for the returns or innovations.
On this basis, a few risk methodologies have been chosen, see the Sec.~\ref{sec:riskMethodologies} for more details.
The figures \ref{fig:tileTestAtDt1_indexes} and \ref{fig:tileTestAtDt1_FX} present the tile test $p$-value as function of the tile length in the $t$ direction for both data sets.
The salient results for both data sets are the following.
First, the historical return methodology is unable to get the short term dynamic, and is rejected for tile lengths up to a few years, or is rejected for all tile lengths.
This is a particularly strong result, and a clear warning for its supporters.
The core reason can be anticipated from our analysis of the UBS stock in the introduction, the tile test places clean statistics on the deficient model for the market dynamics.

Second, the original RiskMetrics methodology and the LM-ARCH + Student innovations have comparable performances, which are not very good.
Other statistics are needed to understand the core reason, but essentially, a fixed distribution is not able to get the peculiarities of each instrument. 
For example, the indexes have positive mean values for the innovations, related to their long term upward trends.
Consequently, the innovations should have a positive mean value, whereas the normal and student distributions used in the present computation are centred.

Third, a methodology using the empirical innovations on a trailing window is very good.
Using the benchmark 1 (left column) shows the clear superiority of this model, while using an adapted benchmark shows that there is still space for improvements against a perfect model.
The volatility forecast seems less important when using an empirical distribution.
The reason lies likely in the structure of the risk forecast based on the historical innovations $\epsilon = r/\sigma$ (more precisely on $\epsilon(t') = r(t')/\tilde{\sigma}(t'-\DT)$) and the risk scenarios for the returns $r = \tilde{\sigma} \epsilon$ (more precisely $\tilde{r}(t)^{(t')} = \tilde{\sigma}(t) \epsilon(t')$ with $t'$ indexing the scenarios). 
Since the algorithm contains both a multiplication and a division by the volatility (at different times), the defects of a volatility model become less important.

\section{Empirical results for $\DT = 10$ business day}
\label{sec:empiricalResults_10d}
\setlength{\figwidth}{0.37\linewidth}
\begin{figure}
	\vspace{-5ex}
	\hspace{0.4\figwidth} \hspace{0.1\figwidth} Benchmark 1 \hfill\hspace{0.3\figwidth} Adapted benchmark \hfill \mbox{}\\[0.5ex]
	\makebox{\parbox[b]{0.4\figwidth}{\raggedright Historical return @1d\vspace*{12ex}}}
	\includegraphics[width=\figwidth]{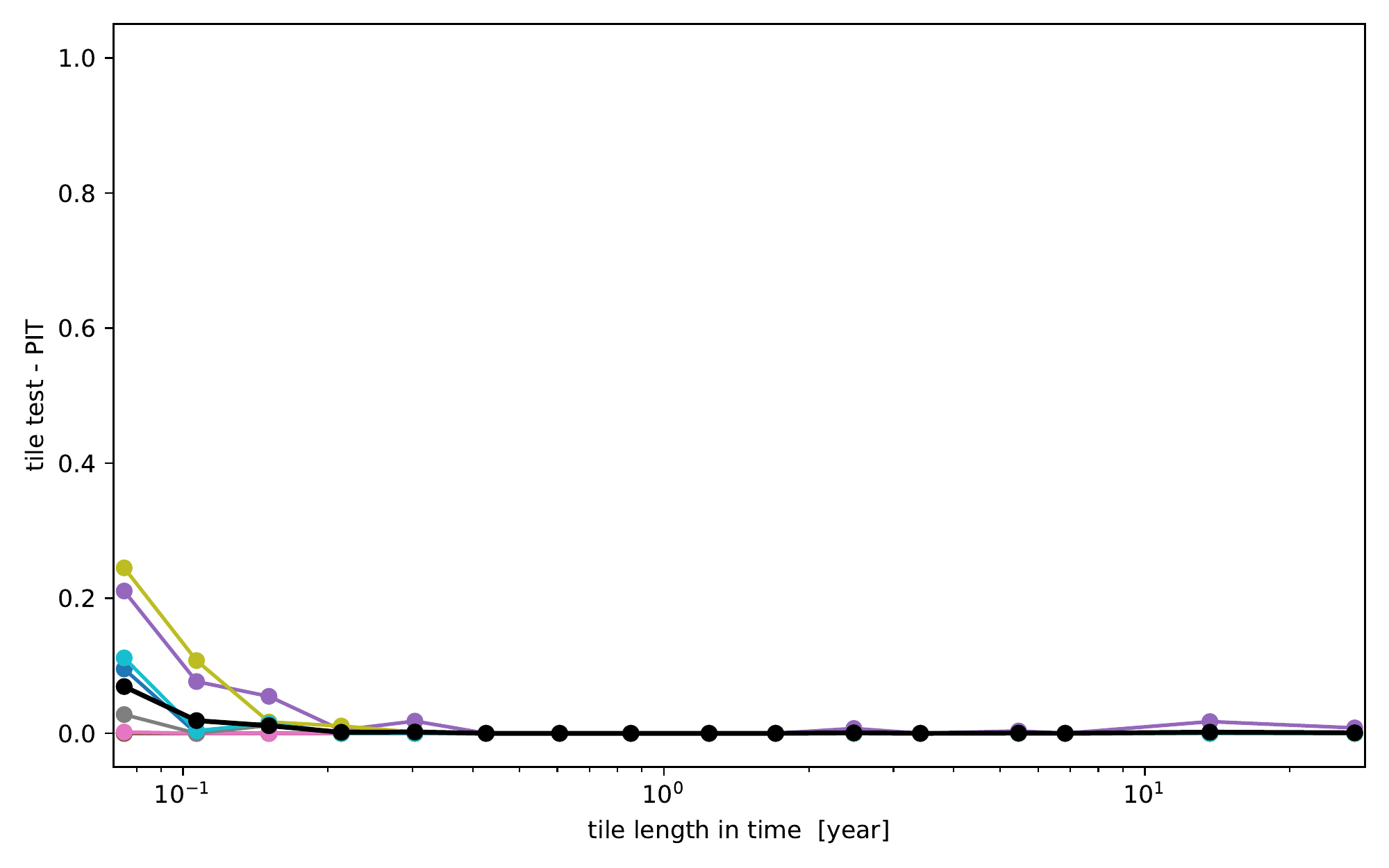}
	\hspace{0.02\linewidth}
	\includegraphics[width=\figwidth]{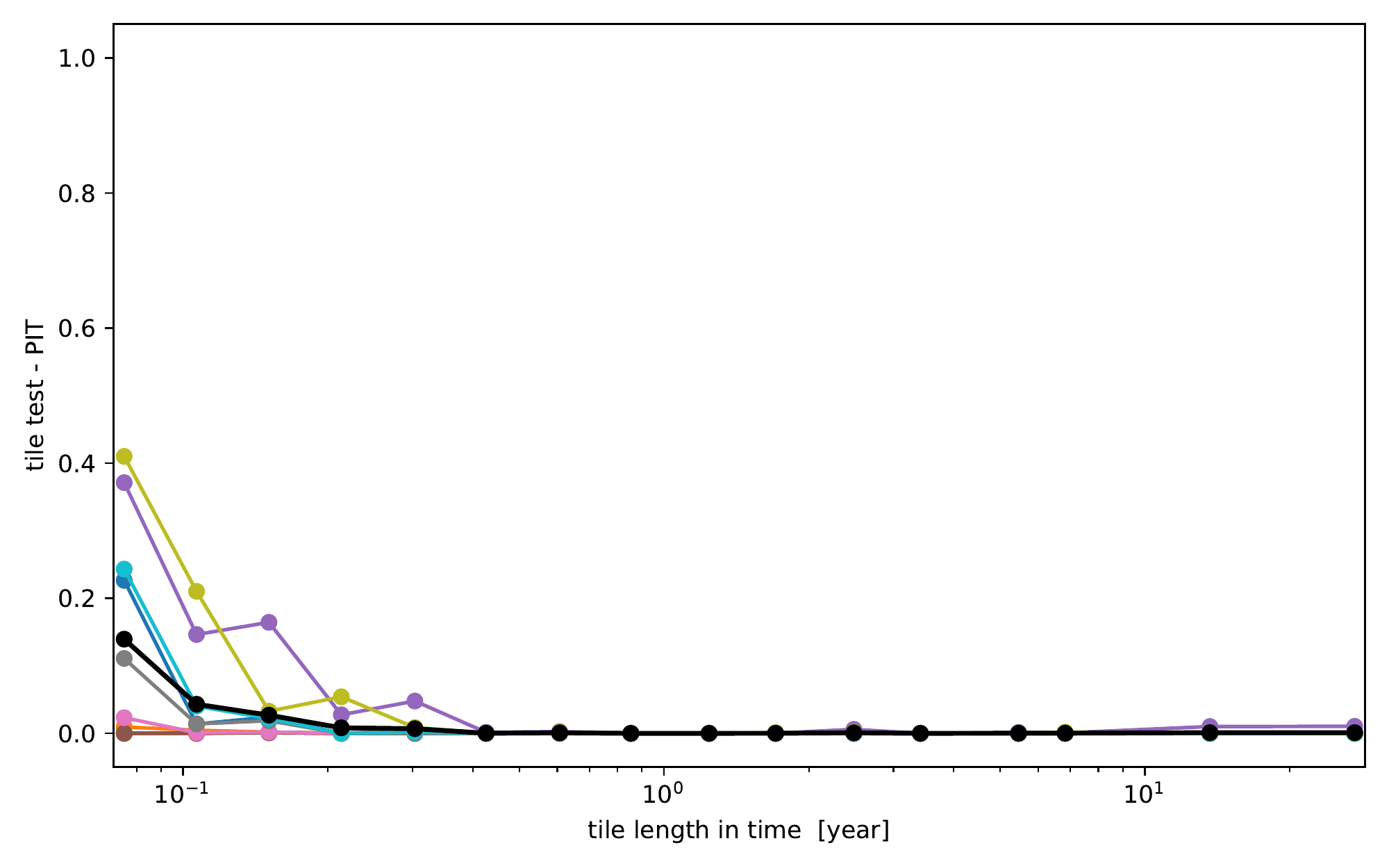}
	\\
	\makebox{\parbox[b]{0.4\figwidth}{\raggedright Historical return @10d\vspace*{12ex}}}
	\includegraphics[width=\figwidth]{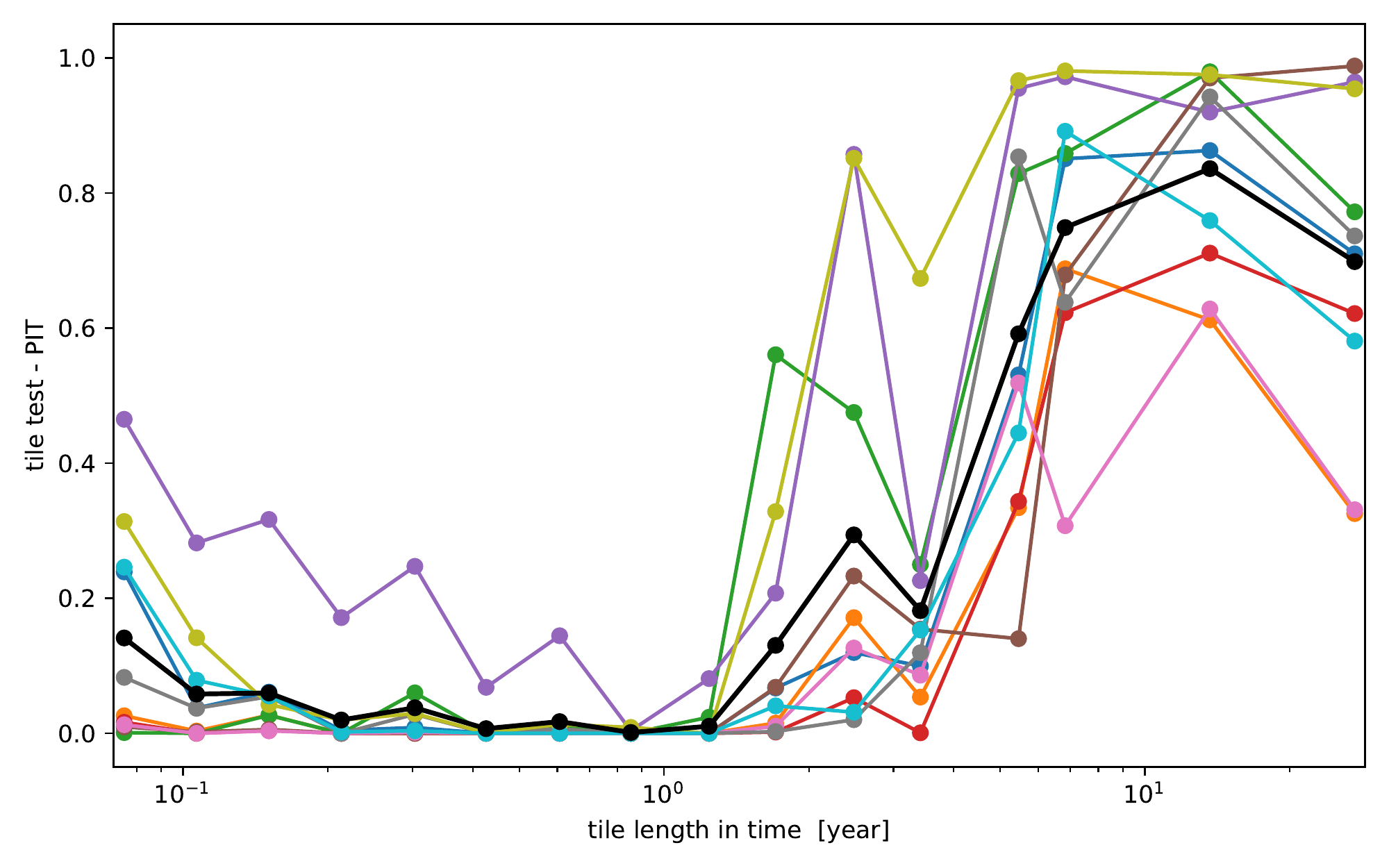}
	\hspace{0.02\linewidth}
	\includegraphics[width=\figwidth]{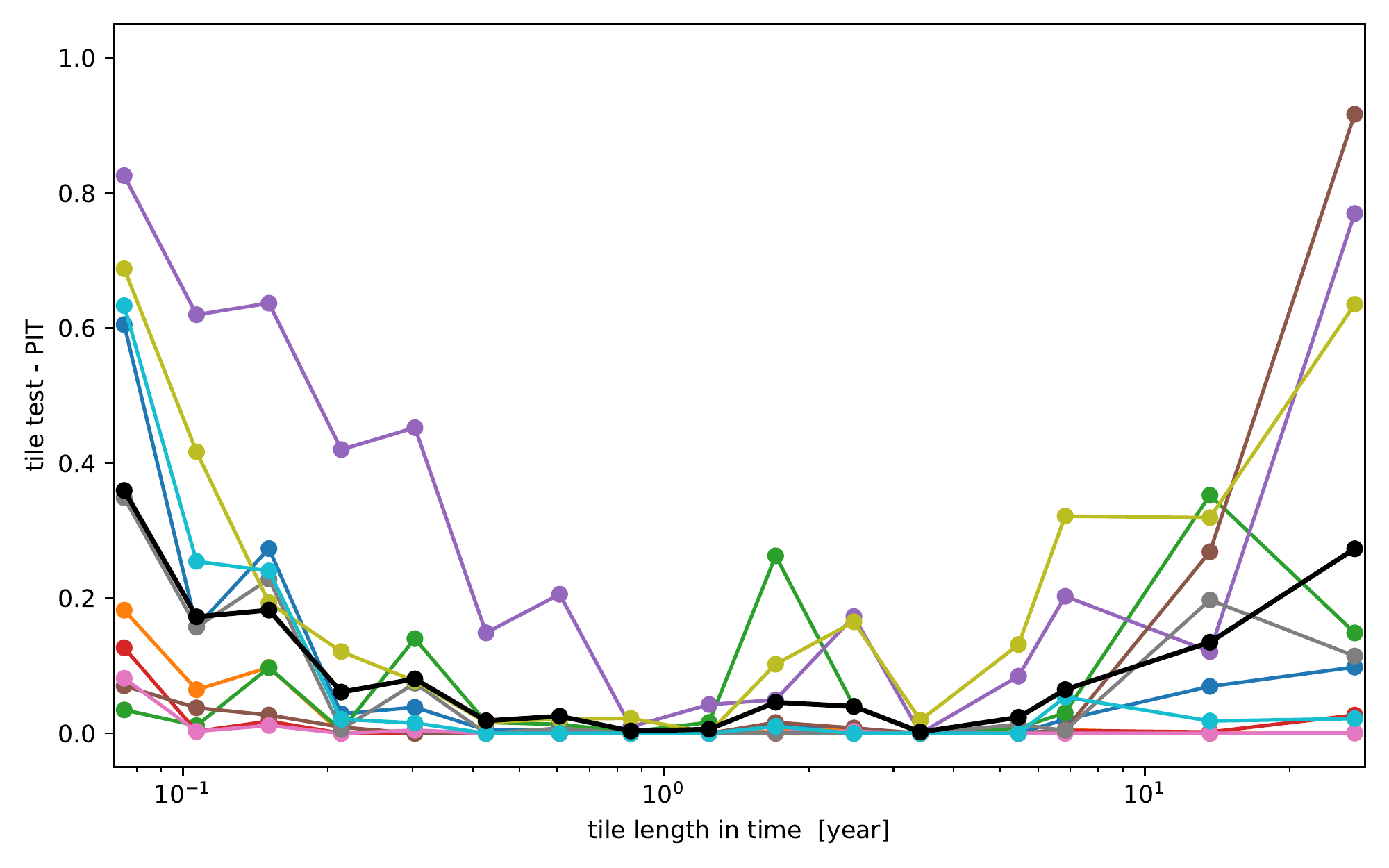}
	\\
	\makebox{\parbox[b]{0.4\figwidth}{\raggedright  RiskMetrics\vspace*{12ex}}}
	\includegraphics[width=\figwidth]{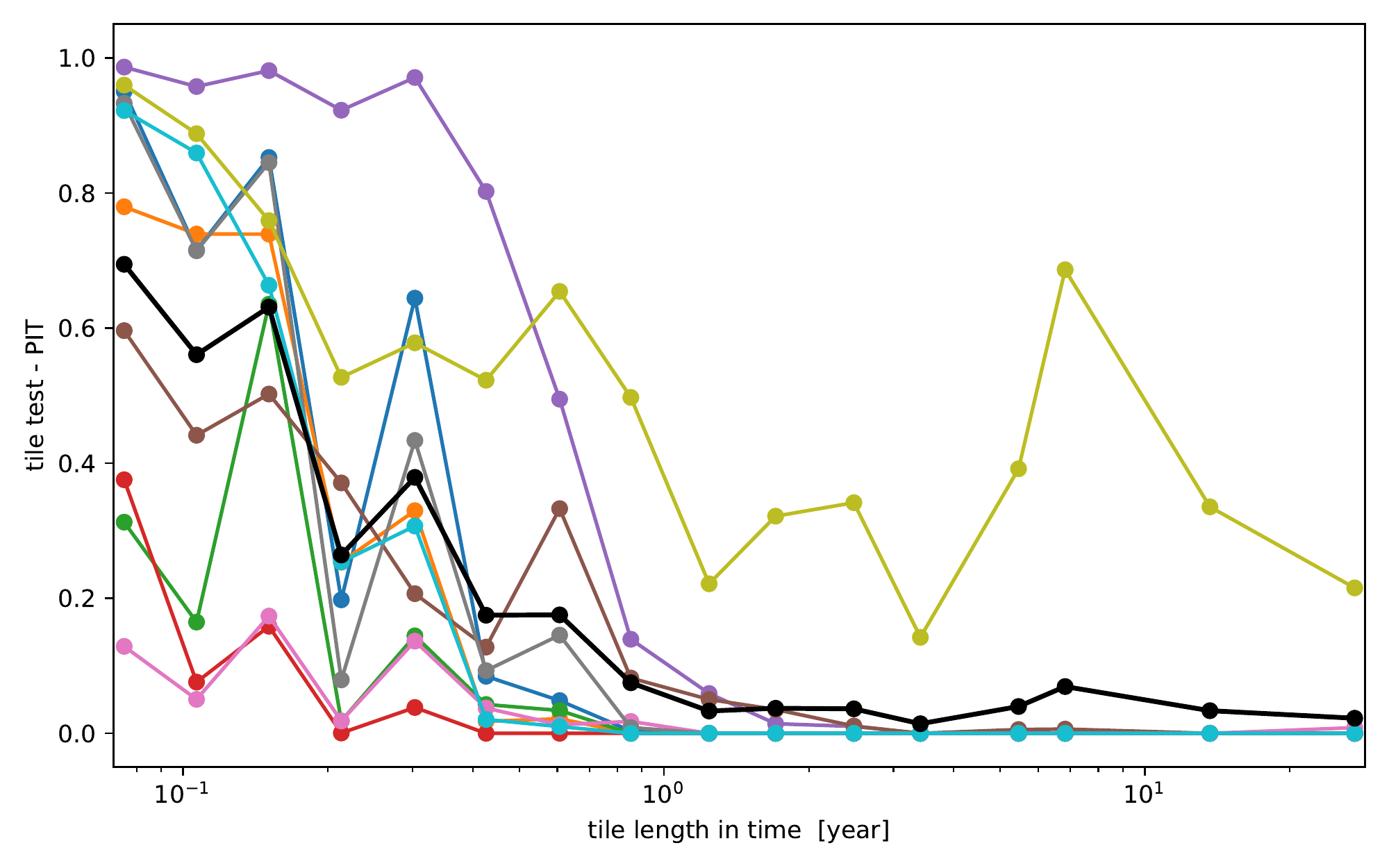}
	\hspace{0.02\linewidth}
	\includegraphics[width=\figwidth]{figures/RiskMetrics/DT_10/tileTest_PIT_bench1_indexes}
	\\
	\makebox{\parbox[b]{0.4\figwidth}{\raggedright  LM-ARCH + Student 6\vspace*{12ex}}}
	\includegraphics[width=\figwidth]{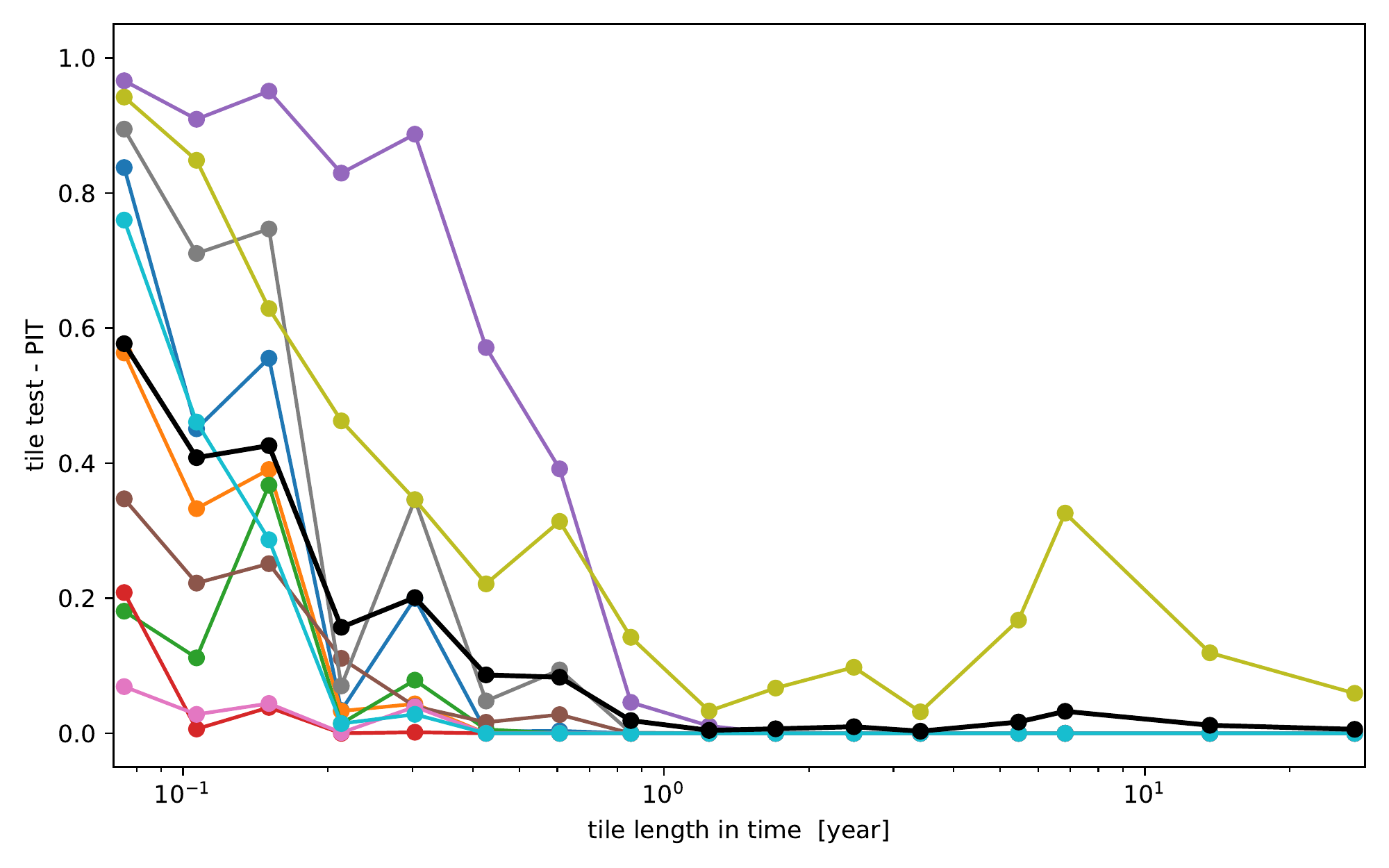}
	\hspace{0.02\linewidth}
	\includegraphics[width=\figwidth]{figures/LMARCH_Student6/DT_10/tileTest_PIT_bench1_indexes}
	\\
	\makebox{\parbox[b]{0.4\figwidth}{\raggedright  LM-ARCH + emp.cdf @1d\vspace*{12ex}}}
	\includegraphics[width=\figwidth]{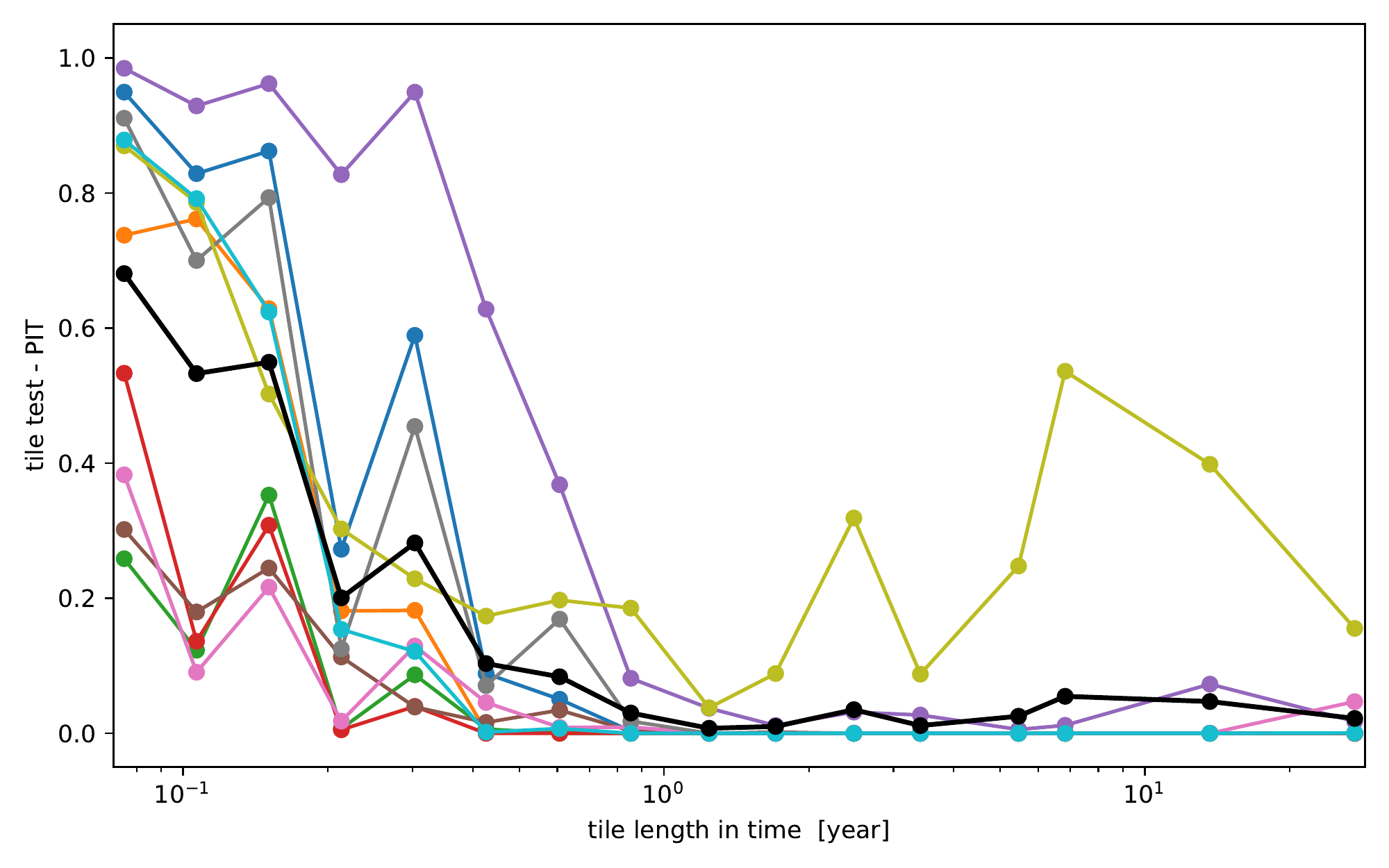}
	\hspace{0.02\linewidth}
	\includegraphics[width=\figwidth]{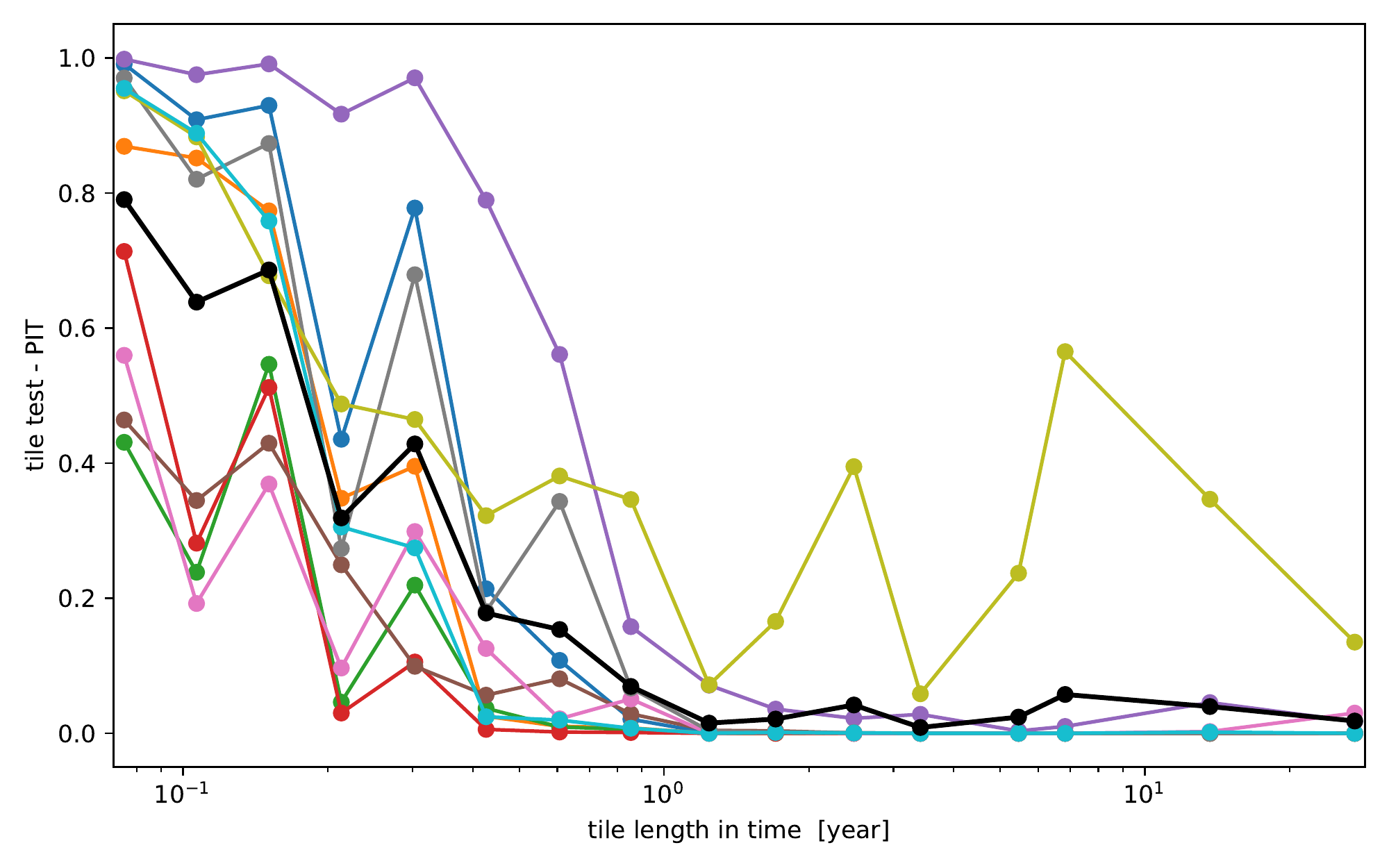}
	\\
	\makebox{\parbox[b]{0.4\figwidth}{\raggedright  LM-ARCH + emp.cdf @10d\vspace*{12ex}}}
	\includegraphics[width=\figwidth]{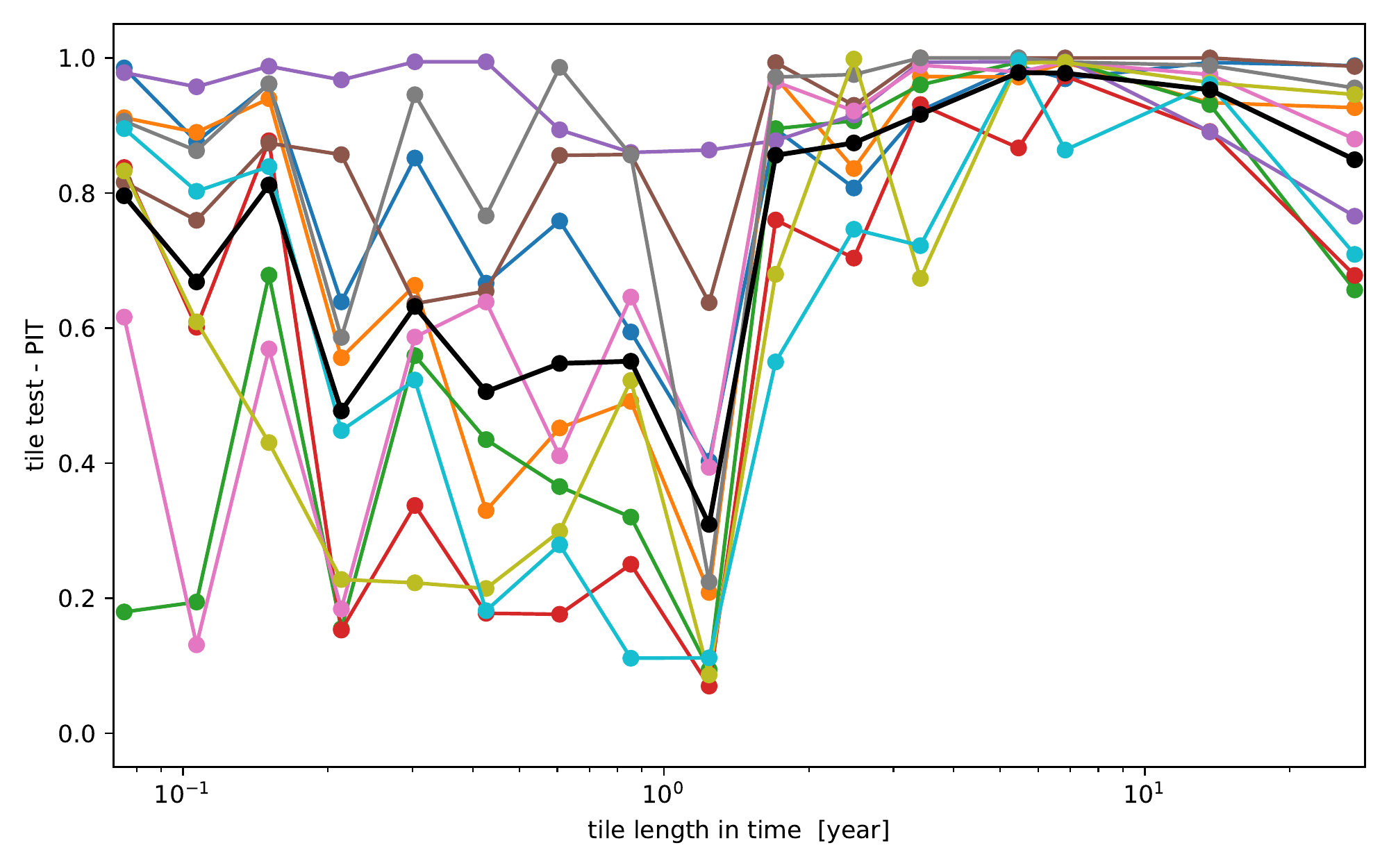}
	\hspace{0.02\linewidth}
	\includegraphics[width=\figwidth]{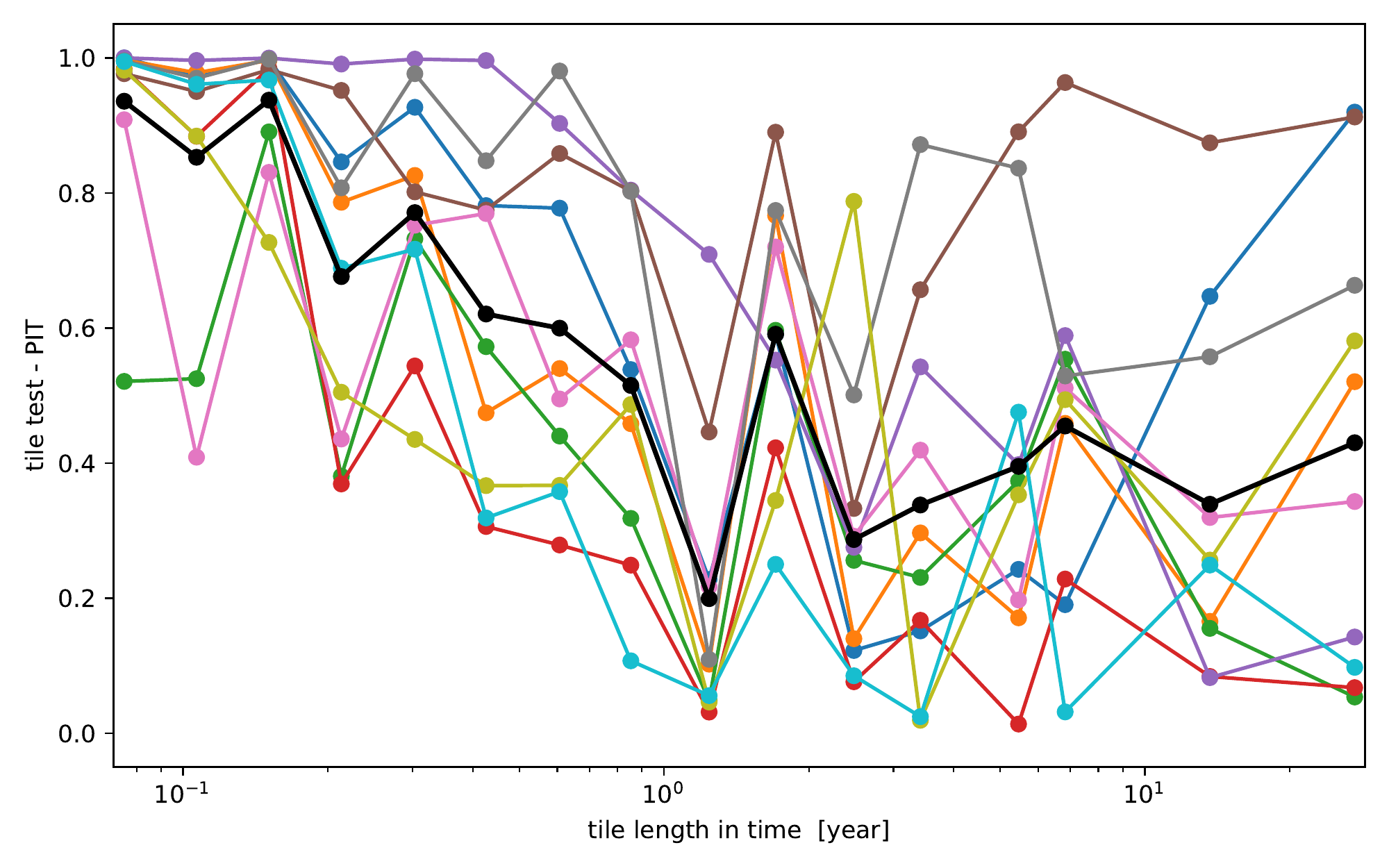}
	
	\caption{Tile test for several risk methodologies, for the 'indexes' sample, at $\DT$ = 10 days. }
	\label{fig:tileTestAtDt10_indexes}
\end{figure}

\begin{figure}
	\vspace{-5ex}
	\hspace{0.4\figwidth} \hspace{0.1\figwidth} Benchmark 1 \hfill\hspace{0.3\figwidth} Adapted benchmark \hfill \mbox{}\\[0.5ex]
	\makebox{\parbox[b]{0.4\figwidth}{\raggedright Historical return @1d\vspace*{12ex}}}
	\includegraphics[width=\figwidth]{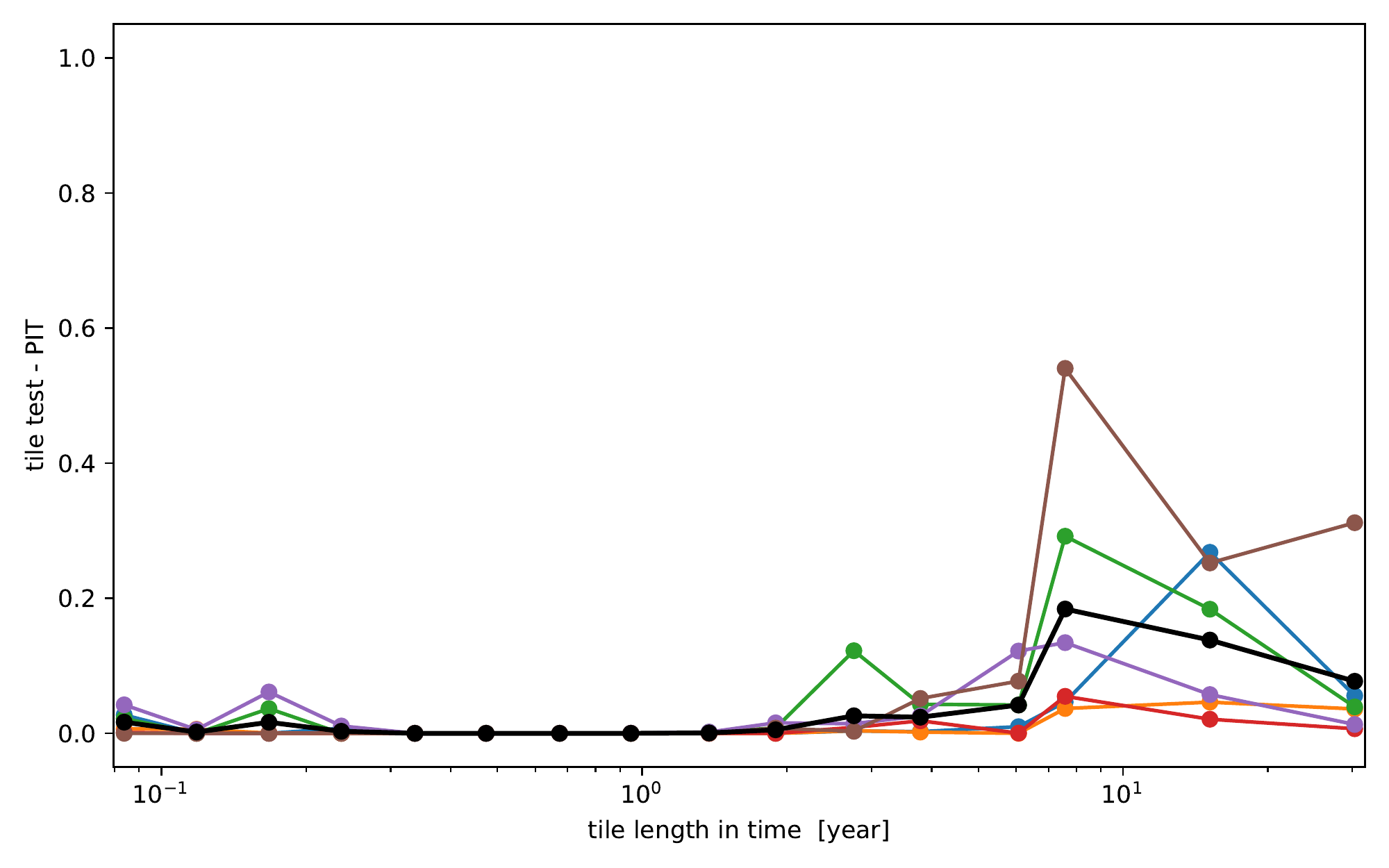}
	\hspace{0.02\linewidth}
	\includegraphics[width=\figwidth]{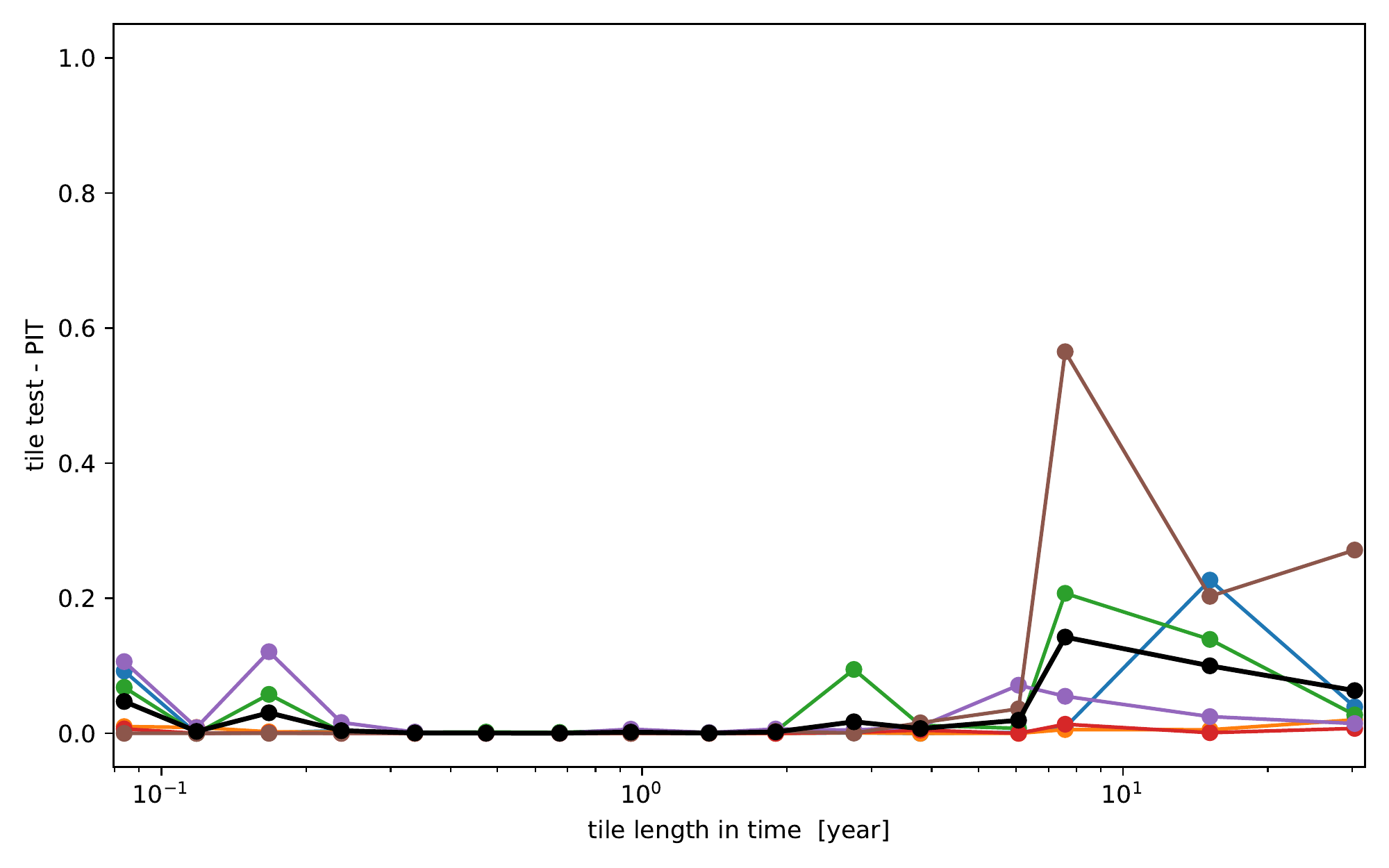}
	\\
	\makebox{\parbox[b]{0.4\figwidth}{\raggedright Historical return @10d\vspace*{12ex}}}
	\includegraphics[width=\figwidth]{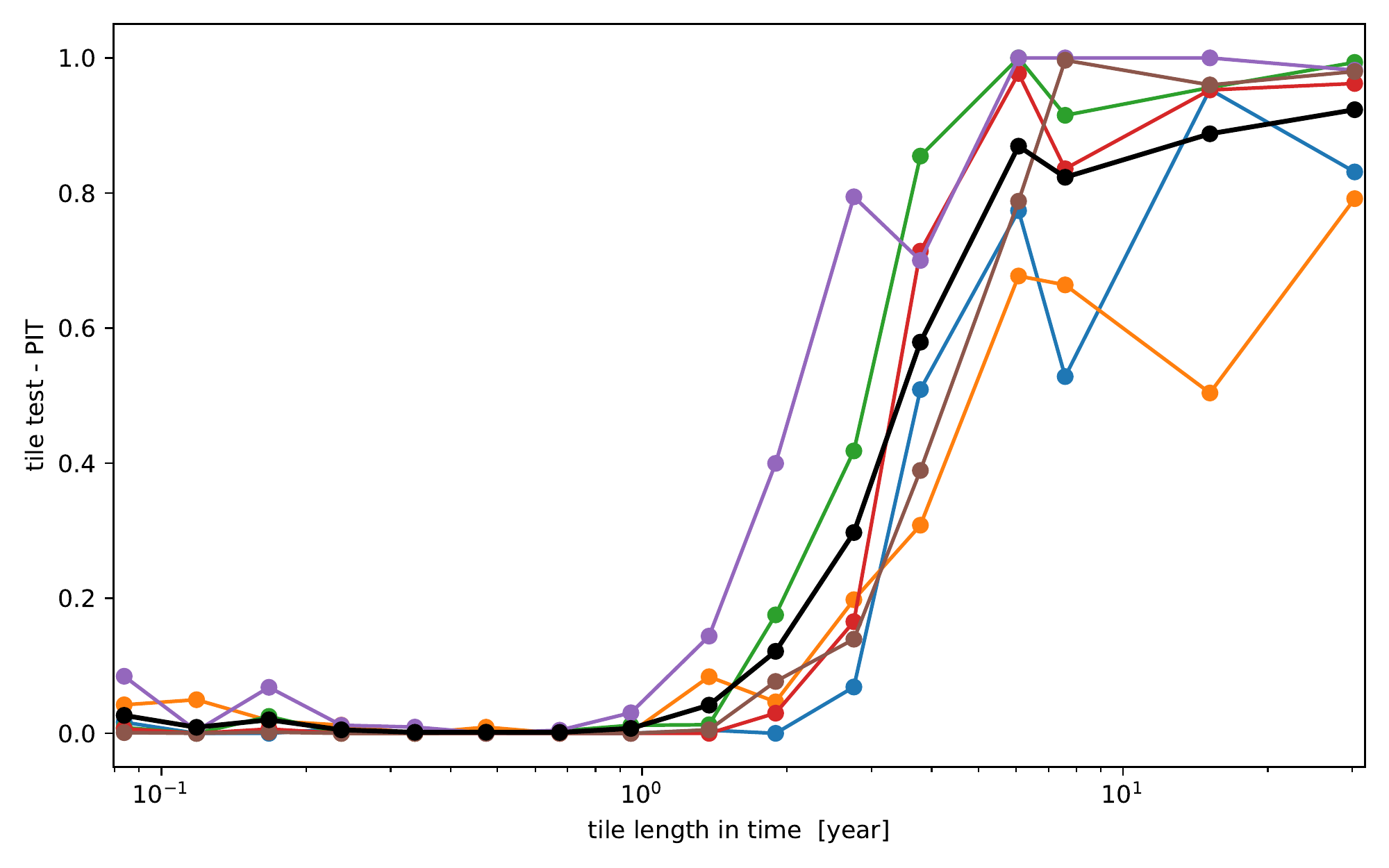}
	\hspace{0.02\linewidth}
	\includegraphics[width=\figwidth]{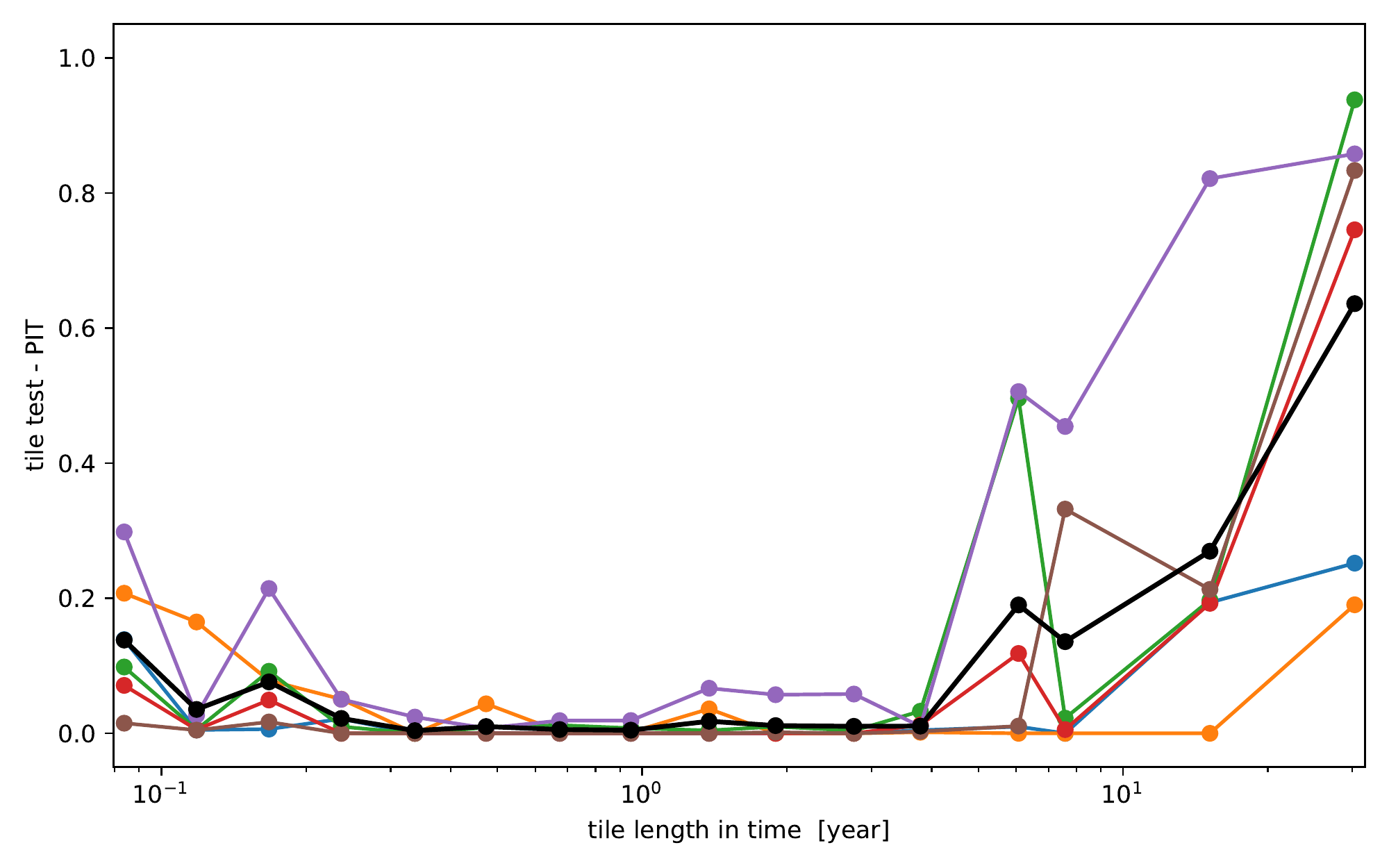}
	\\
	\makebox{\parbox[b]{0.4\figwidth}{\raggedright  RiskMetrics\vspace*{12ex}}}
	\includegraphics[width=\figwidth]{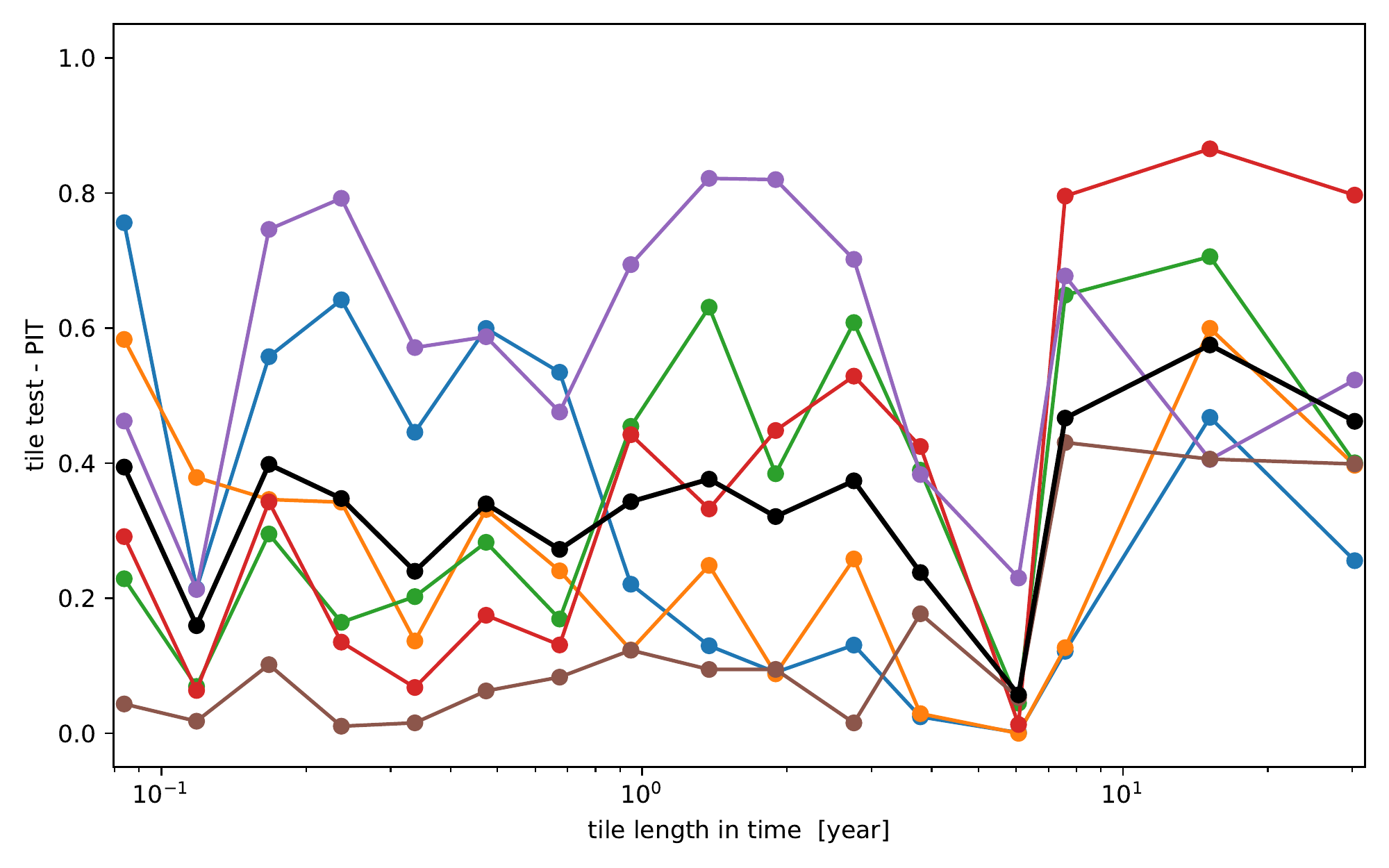}
	\hspace{0.02\linewidth}
	\includegraphics[width=\figwidth]{figures/RiskMetrics/DT_10/tileTest_PIT_bench1_FX}
	\\
	\makebox{\parbox[b]{0.4\figwidth}{\raggedright  LM-ARCH + Student 6\vspace*{12ex}}}
	\includegraphics[width=\figwidth]{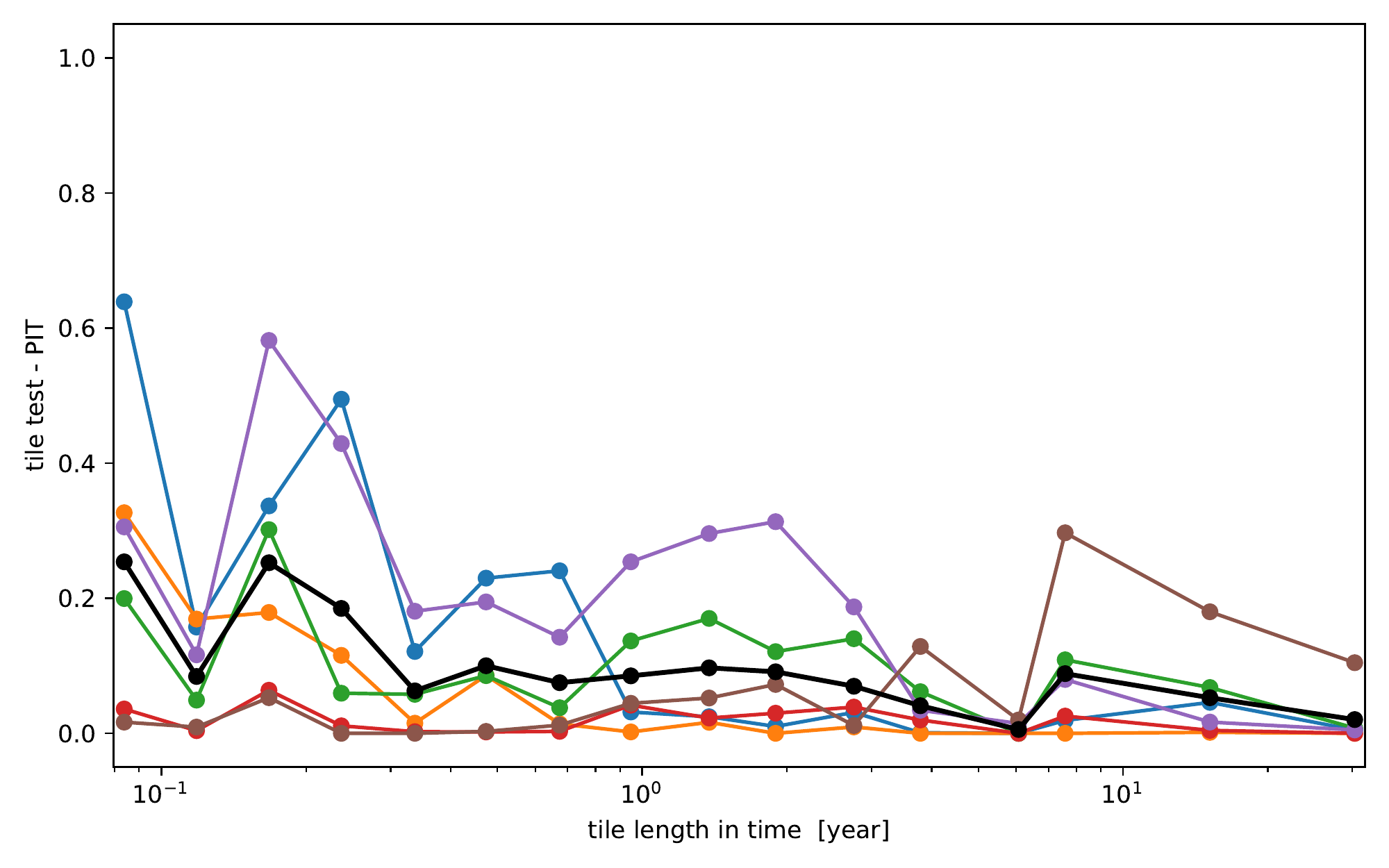}
	\hspace{0.02\linewidth}
	\includegraphics[width=\figwidth]{figures/LMARCH_Student6/DT_10/tileTest_PIT_bench1_FX}
	\\
	\makebox{\parbox[b]{0.4\figwidth}{\raggedright  LM-ARCH + emp.cdf @1d\vspace*{12ex}}}
	\includegraphics[width=\figwidth]{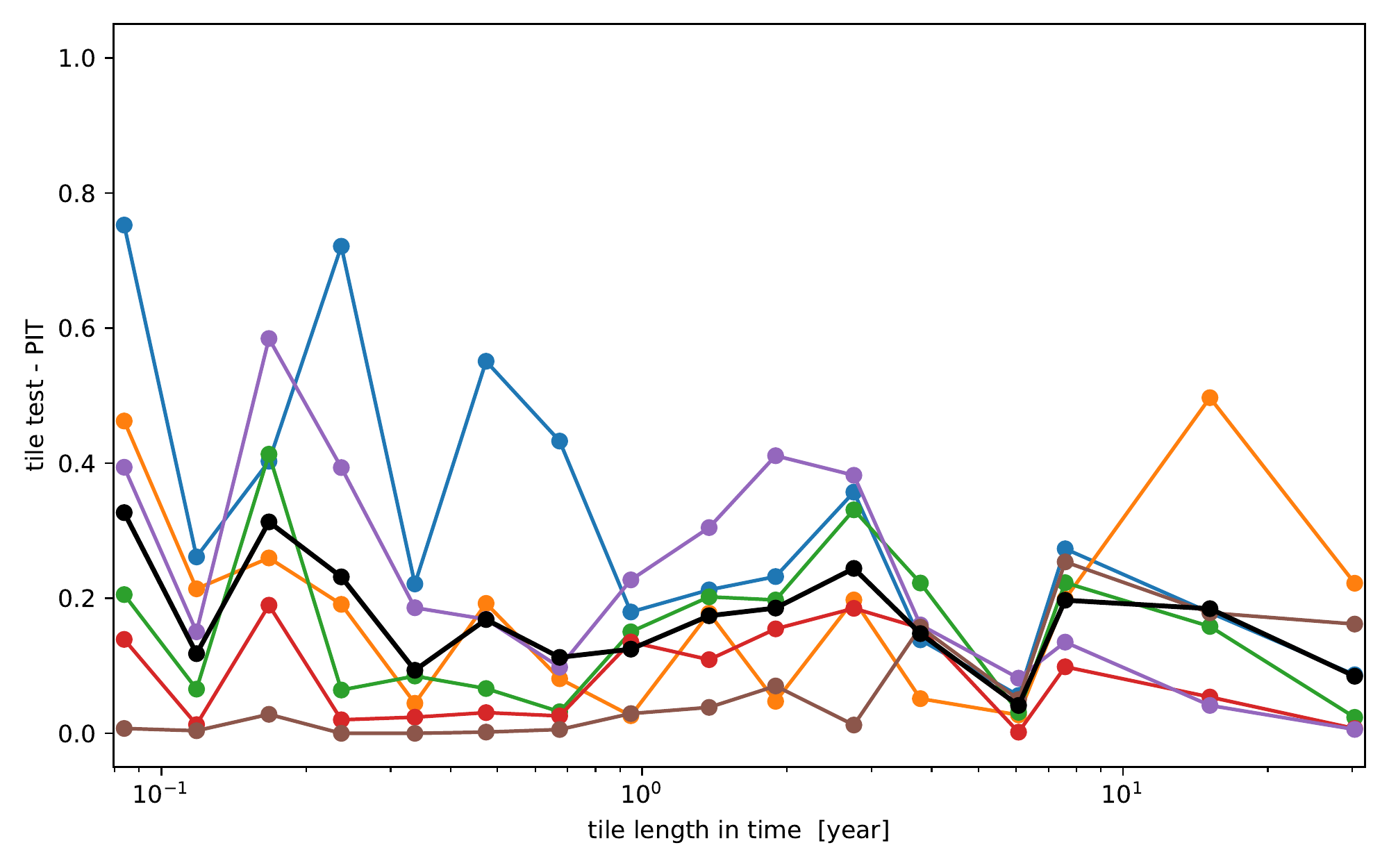}
	\hspace{0.02\linewidth}
	\includegraphics[width=\figwidth]{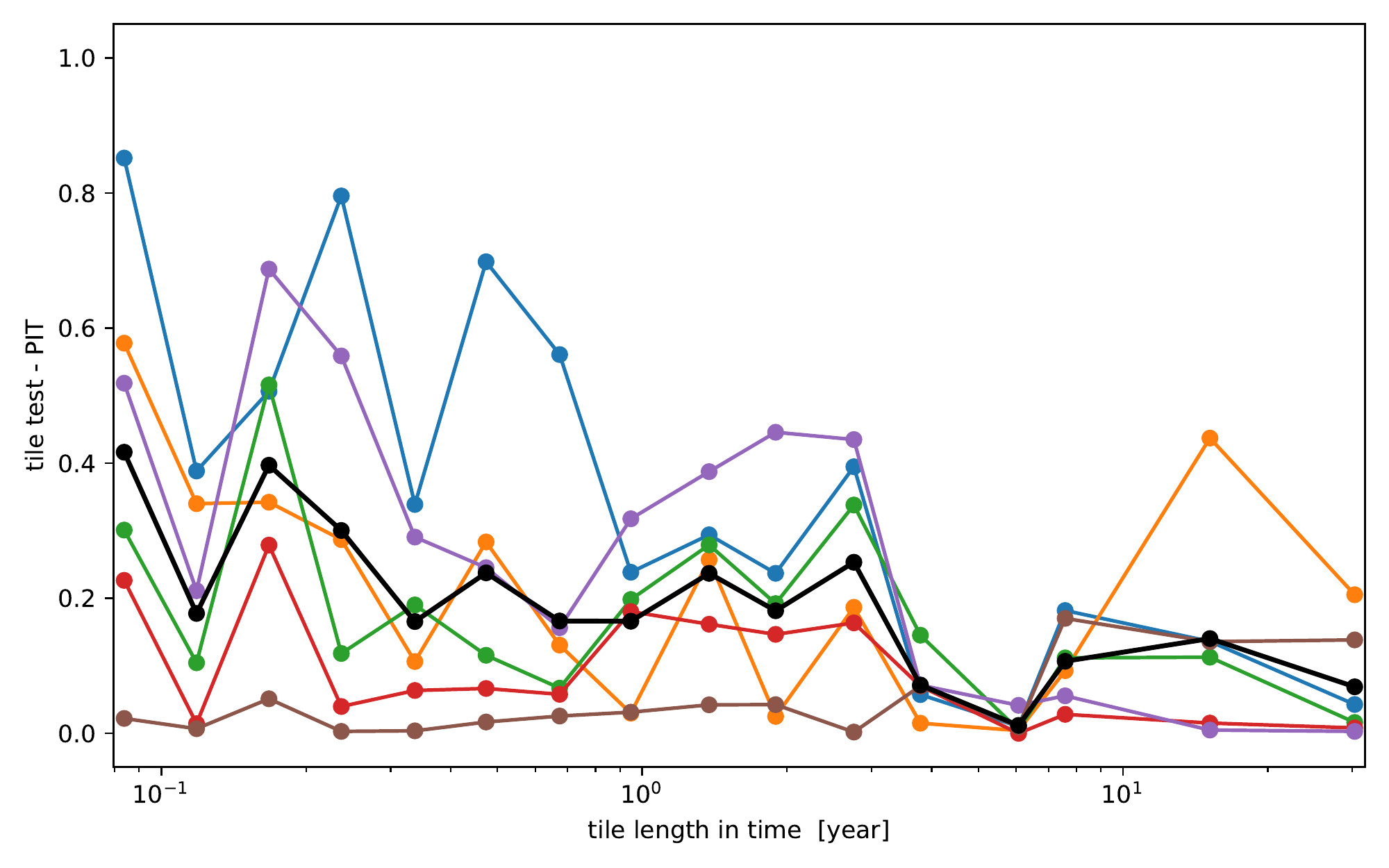}
	\\
	\makebox{\parbox[b]{0.4\figwidth}{\raggedright  LM-ARCH + emp.cdf @10d\vspace*{12ex}}}
	\includegraphics[width=\figwidth]{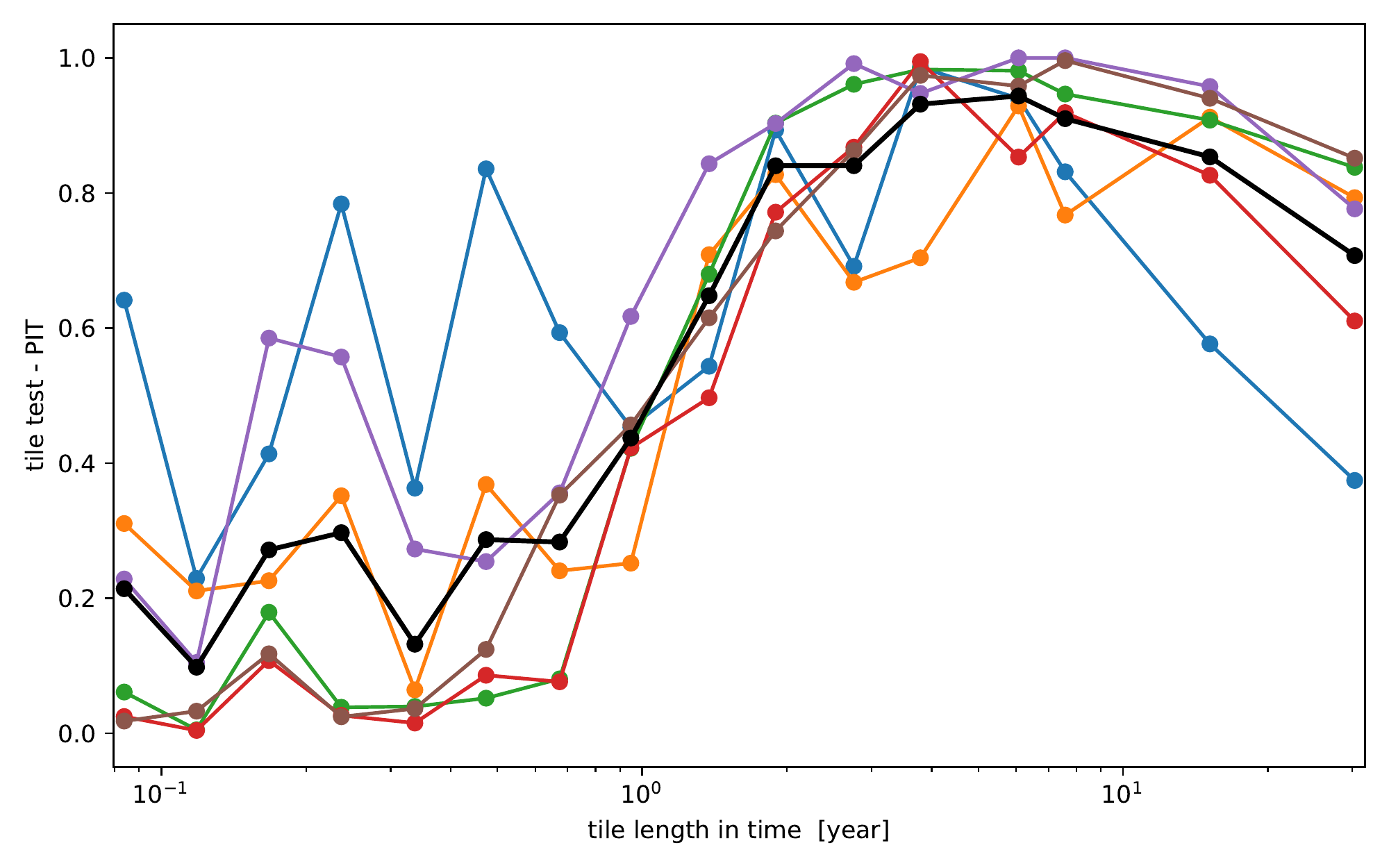}
	\hspace{0.02\linewidth}
	\includegraphics[width=\figwidth]{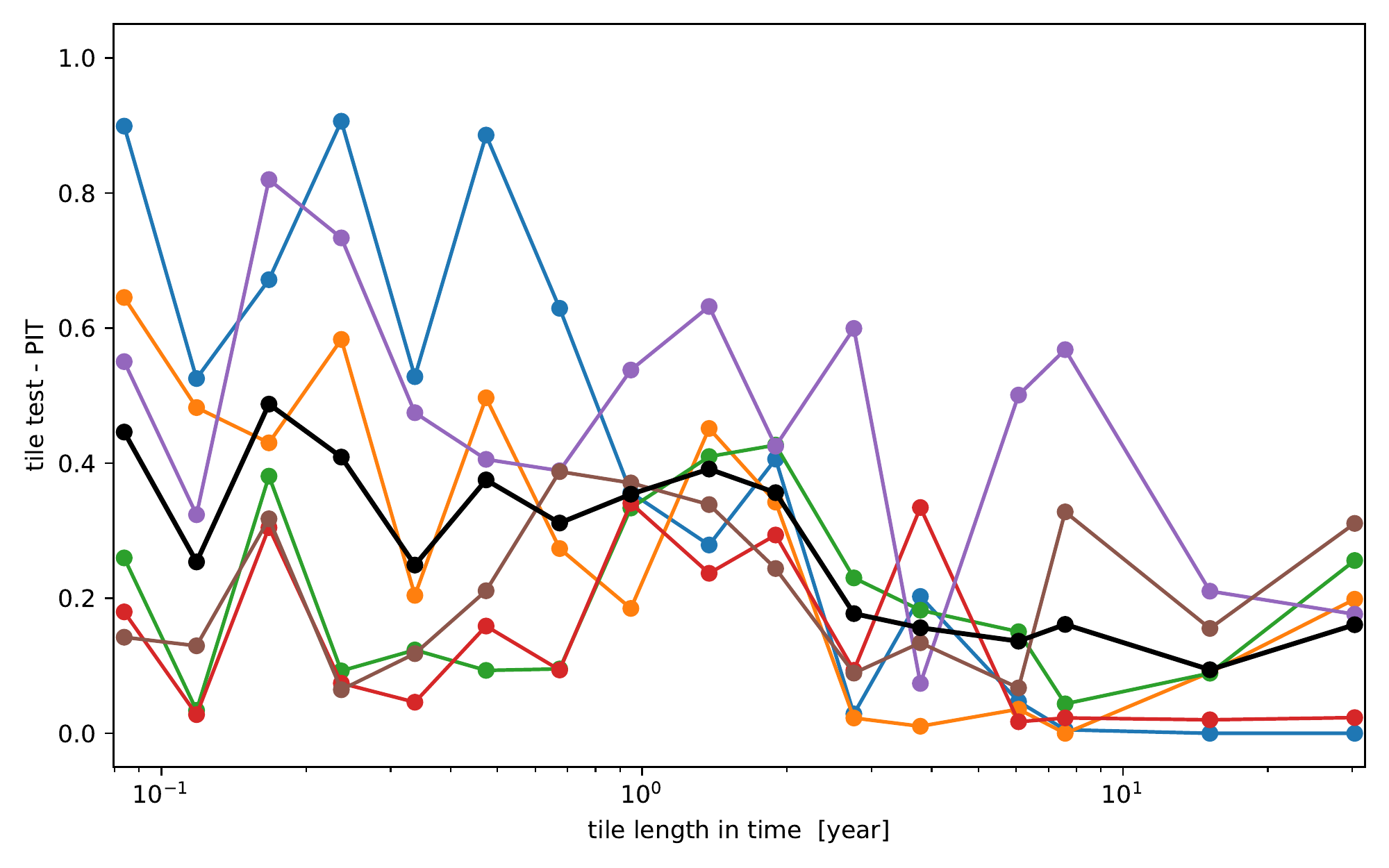}
	
	\caption{Tile test for several risk methodologies, for the 'FX' sample, at $\DT$ = 10 days. }
	\label{fig:tileTestAtDt10_FX}
\end{figure}
In the back-test literature, the results beyond one day are very scarce. 
The reason is the shrinking samples, with the effective sizes going down as $1/\DT$, and an effective test size growing as $\sqrt{\DT}$.
Notice that in our implementation, the computations are always done with the full sample, and the Monte Carlo benchmark have an increasing size, albeit growing as a smaller pace than a square root (see Sec.~\ref{sec:RandomWalkBenchmarks_Properties}). 
Interestingly, the original BIS requirements is to back-test risk evaluations at the 95\% level, for a 1 and 10 days risk horizon. 

In order to tackle this challenge, both very long samples and powerful statistical tests are needed.
In this section, we analyse the same methodologies, but for a risk horizon $\DT =$ 10 business days.
Such validations are very important for practical applications with medium to low turnover, and with an investment horizon ranging from a few weeks to several months.

Two methodologies are added to the panel. 
When computing a historical sample of returns, the time interval to compute the returns must be chosen.
Two choices are natural, namely to compute returns at 1 day or at the risk horizon $\DT$.
Both choices have advantages, namely returns at 1 day lead to the largest sample of independent values, while returns at the risk horizon incorporate possible effects beyond a simple random walk.
At priori, it is not clear which argument is better between the statistical sample and the financial time series model.
This argument can be used for the returns and for the innovations, hence the 2 methodologies added to the panel.

The results are given in figures~\ref{fig:tileTestAtDt10_indexes} and \ref{fig:tileTestAtDt10_FX} for the same empirical samples.
Overall, the results are consistent with the analysis done for a 1 day risk horizon, up to an overall loss of statistical power due to the shrinking samples.
In particular, the historical return methodologies cannot get correctly the short term dynamic, the methodologies with a fixed (centred) distribution have difficulties capturing the peculiarities of time series, and the empirical innovations show the best performances.
On the question raised in the previous paragraph, the analysis shows unambiguously that using innovations (and returns) at the risk horizon is better (in particular for indexes).
This clear-cut answer points to possible specificities of the time series model beyond the heteroskedasticity using an EMA or a LM-ARCH volatility model.

A similar analysis has been done on a set of commodity indexes (agricultural, corn, soybeans, energy, non-precious metal, gold...) and on a panel of Swiss equity (including large, medium and small caps).
All the results are consistent with the above graphs and analysis, showing the robustness of our conclusions.

\FloatBarrier
\section{Conclusions}
We propose a novel statistical test for the back-test of risk methodologies.
It is based on the fundamental object that a risk evaluation produces, namely a forecast for the probability distribution for the expected returns (or losses), and the properties that a correct forecast should have, namely the probability integral transform of the realized returns should be iid with a uniform distribution.
Since these properties should be valid for any sub-sample, a test of uniformity on a tiling allows to probe a risk methodology at various scales. 
Because the dynamics of a risk methodology is a crucial feature to capture risk correctly, we choose a tiling with an  increasingly fine divisions in the $t$ direction, and a fixed number of tiles in the probtile $z$ direction.
This tiling allows to probe a risk computation from the short term dynamic (short tiles, on the left of the graphs) to the asymptotic long term behaviour (long tiles, on the right of the graphs).

A first application of this test using a uniform benchmark leads to a paradox, that is a risk methodology applied on real financial data shows better uniformity that a uniform distribution. 
This paradox is resolved by realizing that a risk computation based on historical data induces negative correlations, creating a ``return to uniformity'' for the probtiles.
Consequently, many risk methodologies based on trailing distributions have too good statistical results when gauged with the natural benchmark.

In order to take this effect into account, a benchmark based on random walks with constant volatility and normal returns is used.
On the random paths, the algorithms used for the risk forecast is applied, possibly including a trailing sample of returns.
This strategy leads to three benchmarks, whether the return distribution is fixed (benchmark 1), is based on a trailing sample of daily returns (benchmark 2), or is using a trailing sample of returns at the risk horizon (benchmark 3).
The quantitative surprise is that the ``return to uniformity''  effect can be strong, leading potentially to distribution with disjoint supports for the benchmarks.
As a side benefit, the benchmarks can be used for any risk horizon and with the full sample, without the need to decimate the sample by a factor $\DT$.
A censoring can also be used for stocks, in order to eliminate spurious effect due to low liquidity.
The benchmark evaluations are purely based on Monte Carlo simulations, and unfortunately an analytical approach seems quite difficult.

Equipped with a powerful test and with a good understanding of the benchmarks, a panel of typical risk methodologies can be investigated. 
The key results are that methodologies based on historical returns do not behave correctly at scales up to a few years, since they do not capture correctly the multi-scales dynamic of the financial markets.
Methodologies based on innovations perform better, with a clear advantage to the methodologies using an empirical distribution for the innovations at the risk horizon.
The figures presented in the paper are based on two data set and two risk horizons, but the empirical results are consistent with other data sets and other risk horizons (1, 2, 5, 10 and 30 days). 

So far, most statistical tests applied on risk evaluation are based on the asymptotic distribution of the returns, the simplest one being counting the number of exceedances at a given probability $\alpha$.
The test proposed in this paper is a significant improvement over the existing tests, in particular since it is a joint test of the dynamics and the distribution.
On the down-side, the test does not provide for a diagnostic of what is right or wrong in a methodology, and other diagnostics should be used in parallel.
In particular, the asymptotic distributions of the returns, probtiles and innovations, as well as lagged correlations for these quantities, give good complementary diagnostics to the tile test.
Finally, the tile test is only based on a forecast in the form of a probability distribution. 
Hence, it can be applied to other fields, say for example weather forecast.

\textbf{Acknowledgement:} The author thanks Samuel Quinodoz and Gwenol Grandperrin for their help with the software development part of this project. 

\bibliographystyle{plainnat}
\bibliography{bibliography}

\end{document}